\documentclass[aps, pra, reprint, superscriptaddress,longbibliography]{revtex4-2}
\usepackage{amsmath}
\usepackage{amsmath}   % For advanced math typesetting
\usepackage{graphicx}  % For including graphics
\usepackage{hyperref}  % For hypertext links
\usepackage{lineno}    % For line numbers
\newcommand{\var}{\mathrm{Var}}  % Define \var command to typeset 'Var' in upright font

\usepackage{lineno,hyperref}
\usepackage{eurosym}
\usepackage{amsmath,amssymb,graphicx}
\usepackage{color}
\usepackage{subcaption}
\modulolinenumbers[5]
\newcommand{\bitem}{\begin{itemize}}
\newcommand{\fitem}{\end{itemize}}
\newcommand{\beq}{\begin{equation}}
\newcommand{\eeq}{\end{equation}}
\newcommand{\beqa}{\begin{eqnarray}}
\newcommand{\eeqa}{\end{eqnarray}}

\begin{document}

\title{Exploring the role of criticality in the quantum Otto cycle fueled by the anisotropic quantum Rabi-Stark model}

\author{He-Guang Xu}

\affiliation{School of Physics, Dalian University of Technology, 116024 Dalian,  China}
\author{Jiasen Jin}
\email{jsjin@dlut.edu.cn}
\affiliation{School of Physics, Dalian University of Technology, 116024 Dalian,  China}
\author{Norton G. de Almeida}
\address{Instituto de Física, Universidade Federal de Goiás, 74.001-970, Goiânia
- GO, Brazil}
\author{G. D. de Moraes Neto}
\email{gdmneto@zjnu.edu.cn}
\affiliation{Department of Physics, Zhejiang Normal University, Jinhua 321004, Zhejiang, P. R. China}

%\date{\today}

\begin{abstract}
 %Quantum heat engines represent a frontier in the field of quantum thermodynamics, offering a novel approach to harnessing and converting heat energy into useful mechanical work. Building upon the principles of quantum mechanics, these engines hold the promise of achieving higher efficiencies and performance compared to their classical counterparts, aiming to revolutionize the way we extract energy from heat sources and offering potential advancements in various applications, including renewable energy technologies and quantum computing.
 %In this paper, we study a quantum Otto engine both in the ideal and finite time cases, with a working substance composed of a two-level system (TLS) interacting with a harmonic oscillator as described by the anisotropic Rabi-Stark model (ARSM), which is notable for presenting both first-order and continuous Quantum Phase Transitions.

Quantum heat machines, encompassing heat engines, refrigerators, heaters, and accelerators, represent the forefront of quantum thermodynamics, offering a novel paradigm for converting heat energy into useful mechanical work. Leveraging quantum mechanical principles, these machines promise superior efficiency and performance compared to classical counterparts, with potential applications in renewable energy and quantum computing. This paper investigates a quantum Otto engine operating in both ideal and finite-time scenarios, employing a two-level system interacting with a harmonic oscillator within the framework of the anisotropic quantum Rabi-Stark model (AQRSM) as the working medium. This model is notable for exhibiting both first-order and continuous quantum phase transitions. By focusing on quantum heat engines, our study reveals that these phase transitions critically modulate the efficiency and power of AQRSM-based engines,  outperforming quantum engines fueled by working medium with harmonic spectrum. Additionally, we explore the impacts of quantum friction and conduct limit cycle analysis in finite-time operations, providing insights into optimizing quantum heat engines for practical implementation.

\end{abstract}

%keywords
\pacs{42.50.Ar, 03.65.Yz, 42.50.Pq}

\maketitle

%\textbf{We investigate the nonclassical feature of photons in the open anisotropic Rabi model that the rotating-wave and counterrotating-wave interactionterms have different coupling constants via the two-photon correlation funtion, Nagativity, Quandrature squeezing of bosonic field, Quantum discord,  Wigner function, and Interference-baced measure.....}
\section{Introduction}

Quantum thermodynamics stands as a rapidly advancing domain aimed at comprehending thermodynamic phenomena at the quantum scale \cite{Kosloff2013Quantum,2013Quantum,goold2016d,Sai2016Quantum,2015Perspective,2017The,deffner2019quantum}. This field attracts significant attention from the scientific community not only due to its role in bridging quantum mechanics and thermodynamics but also for its potential in advancing devices that take advantage of the unique properties of quantum systems as their working medium (WM)\cite{quan2007quantum,quan2009quantum,brunner2012virtual,Mukherjee_2021,souza2022collective,camati2019coherence}. 

Quantum heat engines (QHEs) play a pivotal role in probing the thermodynamic characteristics of quantum systems and harnessing work from heat flow between hot and cold reservoirs \cite{maruyama2009colloquium,scully2010quantum,huang2012effects,2005Autonomous,2009Thermal,goswami2013thermodynamics,kieu2004second,allahverdyan2005work,2012Quantum}. They have attracted considerable attention from scientists and engineers due to the growing demand for efficient energy utilization, storage, and transfer through quantum technologies. Experimentally, these engines have been implemented on various platforms, including trapped ion systems \cite{abah2012single,rossnagel2014nanoscale,rossnagel2016single,maslennikov2019quantum}, optomechanics \cite{zhang2014quantum,2014Theory}, ultracold atoms \cite{2012Isolated,brantut2013thermoelectric}, and superconducting circuits \cite{quan2006maxwell,niskanen2007information,pekola2010decoherence,koski2015chip,pekola2016maxwell}.
 
Critical systems, such as those undergoing quantum phase transitions, are particularly fascinating as WM in quantum heat engines due to their unique properties at criticality, which have been extensively studied in recent literature \cite{campisi2016power, Fogarty_2021, 10.1088/1367-2630/ac963b, PhysRevE.96.022143, interaction_chen, PhysRevResearch.2.043247, e24101458, free_evolution,xu2024universal,mukherjee2024quantum}. Near critical points, these systems exhibit heightened fluctuations and correlations, offering opportunities to enhance work extraction or gain better control over thermodynamic processes. This critical behavior introduces intricate dynamics into the system, resulting in novel scaling laws for work output and efficiency. For instance, the divergence in correlation length and time near a quantum critical point profoundly impacts the engine's efficiency. Moreover, investigating critical systems as WMs facilitates the exploration of universal properties of quantum heat engines. Universality suggests that different systems near criticality may demonstrate similar scaling behaviors, thereby broadening the applicability of the results. This concept implies that the performance of quantum heat engines can be predicted based on the critical exponents of the involved phase transitions \cite{PhysRevResearch.2.043247}.

 The cornerstone model for light-matter interaction is the quantum Rabi model (QRM), which involves a two-level atom (qubit) coupled through dipole interaction to a quantized single-mode cavity field  \cite{rabi1936process,rabi1937space,braak2011integrability,braak2016semi}. This model has been extensively explored across various domains, including quantum optics \cite{scully1997quantum}, entanglement generation  \cite{chen2010entanglement}, quantum thermodynamics \cite{altintas2015rabi,barrios2017role,alvarado2018quantum,wang2024critical}, and the study of quantum phase transitions \cite{hwang2015quantum,hwang2016recurrent,liu2017universal,puebla2016excited,hu2023excited}. When different coupling constants are applied to the rotating-wave (RW) and counter-rotating-wave (CRW) interaction terms, the QRM transforms into the anisotropic quantum Rabi model (AQRM)\cite{yu2013goldstone,xie2014anisotropic,tomka2014exceptional}. Additionally, omitting the CRW terms results in the Jaynes-Cummings model (JCM) \cite{jaynes1963comparison}. Notably, both the AQRM and the JCM exhibit first-order quantum phase transitions (QPTs)\cite{xie2014anisotropic,zhu2016effects}. These QPTs significantly impact the quantum thermalization dynamics and the persistence of quantum correlations at thermal equilibrium \cite{fan2020quantum,liu2023quantum,xu2024persisting}.

%==========================================
Moreover, the Quantum Rabi-Stark model (QRSM) is an extension of the QRM that incorporates nonlinear coupling terms\cite{grimsmo2013cavity,grimsmo2014open,eckle2017generalization}. This model was inspired by recent experimental advancements\cite{cong2020selective} and has been extensively studied in theoretical works\cite{maciejewski2015exactly,xie2019quantum,xie2019exact,cong2020selective}. The QRSM has been analyzed using the Bargmann space approach \cite{eckle2017generalization,maciejewski2015exactly} and solved using the Bogoliubov operator approach (BOA)\cite{xie2019quantum}. These studies have revealed many exotic properties of the QRSM, such as first-order QPT and spectral collapse \cite{xie2019quantum}. Furthermore, the AQRM with Stark coupling terms, also known as the anisotropic quantum Rabi-Stark model (AQRSM), has been found to exhibit both first-order and continuous QPTs \cite{xie2020first}.

In this paper, we perform a comprehensive examination of a thermal engine utilizing the AQRSM as the working substance. Our study begins by thoroughly analyzing the ideal scenario of infinite operation time and zero power delivered, focusing on the intricate relationship between phase transitions and their influence on enhancing efficiency, work throughout the quantum Otto cycle. Additionally, we construct a detailed phase diagram to provide insights into the operational regimes, shedding light on the behavior of the coupled system compared to engines operating with a qubit or a quantum harmonic oscillator.

Expanding our investigation beyond the ideal scenario, we explore finite-time operations, where the trade-off between power and efficiency becomes prominent, along with the effects of quantum friction. Here, we highlight the crucial role played by the spectral characteristics of the working medium, including level crossings, degeneracies, and quasi-degenerate points, in shaping the performance of quantum engines. Furthermore, we incorporate a limit cycle analysis to deepen our understanding of the system behavior under finite-time conditions. 

Acknowledgments to previous investigations on light-matter heat engines\cite{altintas2015rabi,barrios2017role,alvarado2018quantum,wang2024critical}, which considered the QRM as the WM, are due. However, these studies did not encompass first-order QPTs, which are absent in the QRM, and focused on the ideal scenario of zero power delivery and extremely low temperatures. Notably, only Ref.\cite{wang2024critical} addressed the second-order QPTs occurring at an infinite frequency ratio for the Stirling cycle.

 The paper is organized as follows: In Section II A, we introduce the Anisotropic Quantum Rabi-Stark model, describe the Quantum Otto Cycle, and present the calculations for work and heat, as well as the operation in finite time. In Section III, we present and discuss the results, focusing on the working operational regimes for the ideal adiabatic process, efficiency, and work. Additionally, for the finite-time cycle, we focus on the Heat engine and perform an analysis of entropy production and how the power and efficiency are affected by the light-matter coupling. In Section IV, we present our conclusions.

%==========================================
\section{THEORETICAL FRAMEWORK}
\subsection{Anisotropic quantum Rabi-Stark model}
The Anisotropic Quantum Rabi-Stark Model (AQRSM) describes the coupling between a two-level system (qubit) and a single-mode bosonic field, experiencing both the Rotating-Wave (RW) and Counter-Rotating-Wave (CRW) interactions asymmetrically. Additionally, it incorporates a nonlinear coupling term between the atomic inversion and the cavity photon number. The Hamiltonian of the AQRSM\cite{xie2020first} is given by ($\hbar = 1$)
\begin{equation}
\begin{split}
    \hat{H} & = \omega\hat{a}^{\dagger}\hat{a}+\frac{\Delta}{2}\hat{\sigma}_{z}+U\hat{a}^{\dagger}\hat{a}\hat{\sigma}_{z}\\
    & + \lambda_{1}(\hat{a}\hat{\sigma}^{+}+\hat{a}^{\dagger}\hat{\sigma}^{-})+\lambda_{2}(\hat{a}^{\dagger}\hat{\sigma}^{+}+\hat{a}\hat{\sigma}^{-}).\label{eq:hamiltonian-aqrm}
\end{split}
\end{equation}
In the above, $\omega$ is the frequency of the bosonic field described by annihilation (creation) operator $\hat{a}$ ($\hat{a}^\dagger$), $\Delta$ is the transition frequency of qubit described  by Pauli matrices $\hat{\sigma}_{x,y,z}, \hat{\sigma}^{\pm}=(\hat{\sigma}_{x}\pm i\hat{\sigma}_{y})/2$ . $U$ denotes the nonlinear Stark coupling strength, while $\lambda_1$ and $\lambda_2$ account for the qubit-boson coupling strengths for the RW and CRW interactions, respectively. One key aspect of the AQRSM in Eq.\eqref{eq:hamiltonian-aqrm} is that it preserves the parity symmetry $\mathbb{Z}_{2}$. This fact can be evidenced by considering the total number of excitations operator $\hat{N}=\hat{a}^\dagger\hat{a}+\hat{\sigma}^+\hat{\sigma}^-$ and the parity operator $\hat{\pi}=\mathrm{exp}\left(i\pi\hat{N}\right)$. The conserved parity  $\mathbb{Z}_{2}$
possesses two eigenvalues, $\left\langle\hat{\pi}\right\rangle = \pm 1$, depending on whether the total number of excitations are even or odd. 

The AQRSM is notable for its manifestation of both first-order and continuous quantum phase transitions (QPT)\cite{xie2020first}. First-order QPTs are marked by abrupt changes in ground-state properties during the transition, especially in the infinite-volume (thermodynamic) limit. In contrast, continuous QPTs exhibit smooth changes in ground-state features, including the emergence of divergent length scales in quantum correlation functions and a gapless energy spectrum in the infinite-volume limit.

The first-order QPT arises in the AQRSM at the energy crossing between the ground state and the first excited state, occurring when $\omega=1$ at

\begin{equation}
\lambda_{1c}=\sqrt{\frac{\Delta(1-U^{2})}{U(1+r^{2})+1-r^{2}}},
\end{equation}
where $r=\lambda_{2}/\lambda_{1}$ represents the degree of anisotropy. This transition signifies a substantial shift in the system's behavior and is crucial for understanding its quantum properties. Continuous QPT, on the other hand, occurs in the QRM at an infinite frequency ratio of $\Delta/\omega$\cite{ hwang2015quantum}. At a critical coupling, the system undergoes a second-order quantum phase transition into a superradiant phase, which has been observed in trapped-ion settings\cite{cai2021observation}, a phenomenon that naturally manifest itself in the AQRSM as well\cite{liu2017universal}.

Interestingly, AQRSM exhibits another form of continuous QPT, in a less restrictive physical realization, at finite frequency ratios $\Delta/\omega$ for the special Stark coupling $|U|=1$. When $U=\pm1$, AQRSM can be mapped into an effective quantum oscillator. The low energy spectrum separates into two branches: the upper one with $E>\mp\frac{\Delta}{2}-2\kappa^{2}\alpha^{2}$, and the lower one with $E< E_{c}^{\pm}=\mp\frac{\Delta}{2}-2\alpha^{2}$, where $\alpha=(\lambda_{1}+\lambda_{2})/2$ and $\kappa=(1-r )/(1+r)\leq1$. The real lower spectra only exist before the critical coupling $\alpha_{c}^{\pm}=\sqrt{\frac{1\mp\Delta\pm\kappa}{2}}$.

At the critical points $\alpha_{c}^{\pm}$, the phenomenon known as spectral collapse occurs, where the discrete low energy levels converge to a shared energy value $E_{c}^{\pm}$, signaling the closure of the energy gap within the system at the critical coupling $\alpha_{c}^{\pm}$. It is therefore suggested that gapless Goldstone mode excitations appear for $\alpha_{c}^{\pm}$. Furthermore, beyond the critical point, the system experiences the breaking of parity symmetry due to the infinite degeneracy of all states, characteristic of a second-order QPT.
We note that the presence of QPT in the AQRSM model belongs to a distinct universality class compared to models that exhibit phase transitions in the infinite frequency limit\cite{chen2020quantum}.

\begin{figure*}[!htbp]
\begin{center}
%\vspace{-2.2cm}
\includegraphics[width=0.9\textwidth]{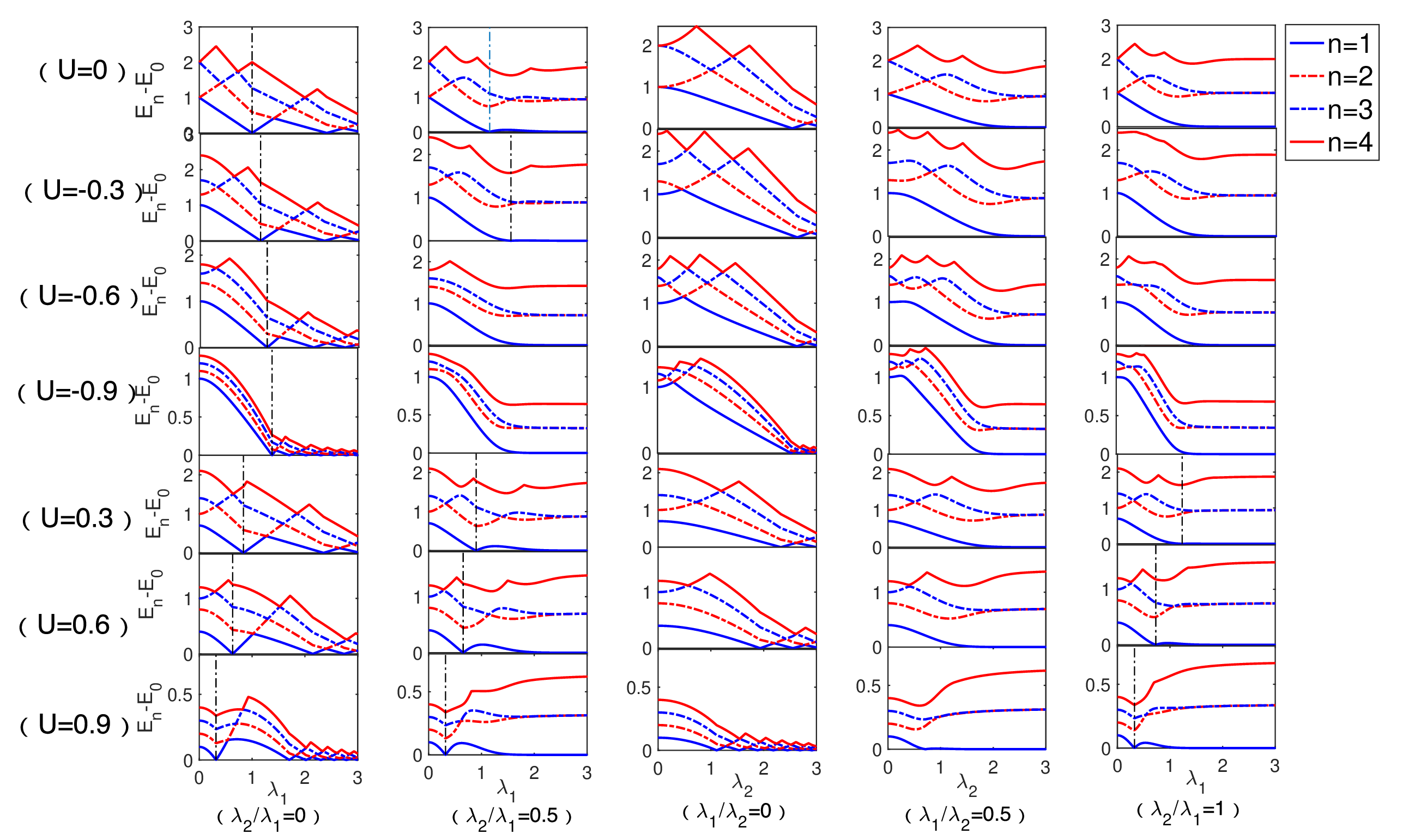}
%\vspace{-1.0cm}
\end{center}
 \caption{Energy spectrum diagram for the anisotropic quantum Rabi-Stark model. The plots show differences in energy levels as functions of the light-matter coupling strengths $\lambda_{1}$ (RW) and $\lambda_{2}$ (CRW) for various anisotropy ratios $\lambda_{2}/\lambda_{1}$ and $\lambda_{1}/\lambda_{2}$. Energy levels are computed over a range of Stark interaction strengths $U$ from -0.9 to 0.9, in increments of 0.3. The vertical dash-dot black lines indicate the position of the first-order phase transition according to Eq.(2), providing insight into the critical behavior of the system. Additional system parameters are set as $\Delta=1$ and $\omega=1$}
\label{spectrum}
\end{figure*}

The spectral characteristics of the working medium, such as level crossings, degeneracies, and quasi-degenerate points, play a crucial role in determining the performance of a quantum engine. These features can either enhance or suppress specific thermodynamic processes, thereby influencing the engine's efficiency and overall performance. In Fig.\ref{spectrum}, we present the energy spectrum of the AQRSM in relation to the ground state energy, considering various parameters such as the light-matter coupling strengths $\lambda_1$ and $\lambda_2$, and the nonlinear Stark interaction $U$. Notably, the spectrum reveals level crossings characterized by changes in ground state parity, leading to infinite first-order quantum phase transitions as the coupling strength increases. Moreover, the presence of non-degenerate and near-degenerate points further impacts the operation and thermodynamic behavior of the engine. Understanding the spectrum of the working medium is vital not only for uncovering the underlying physics behind the emergence of thermal correlations but also for determining the characteristics of non-equilibrium states accessible during the finite-time cycle of the quantum engine.

\subsection{Quantum Otto Cycle}

\begin{figure}[!htbp]
\begin{center}
%\vspace{-2.2cm}
\includegraphics[width=0.5\textwidth]{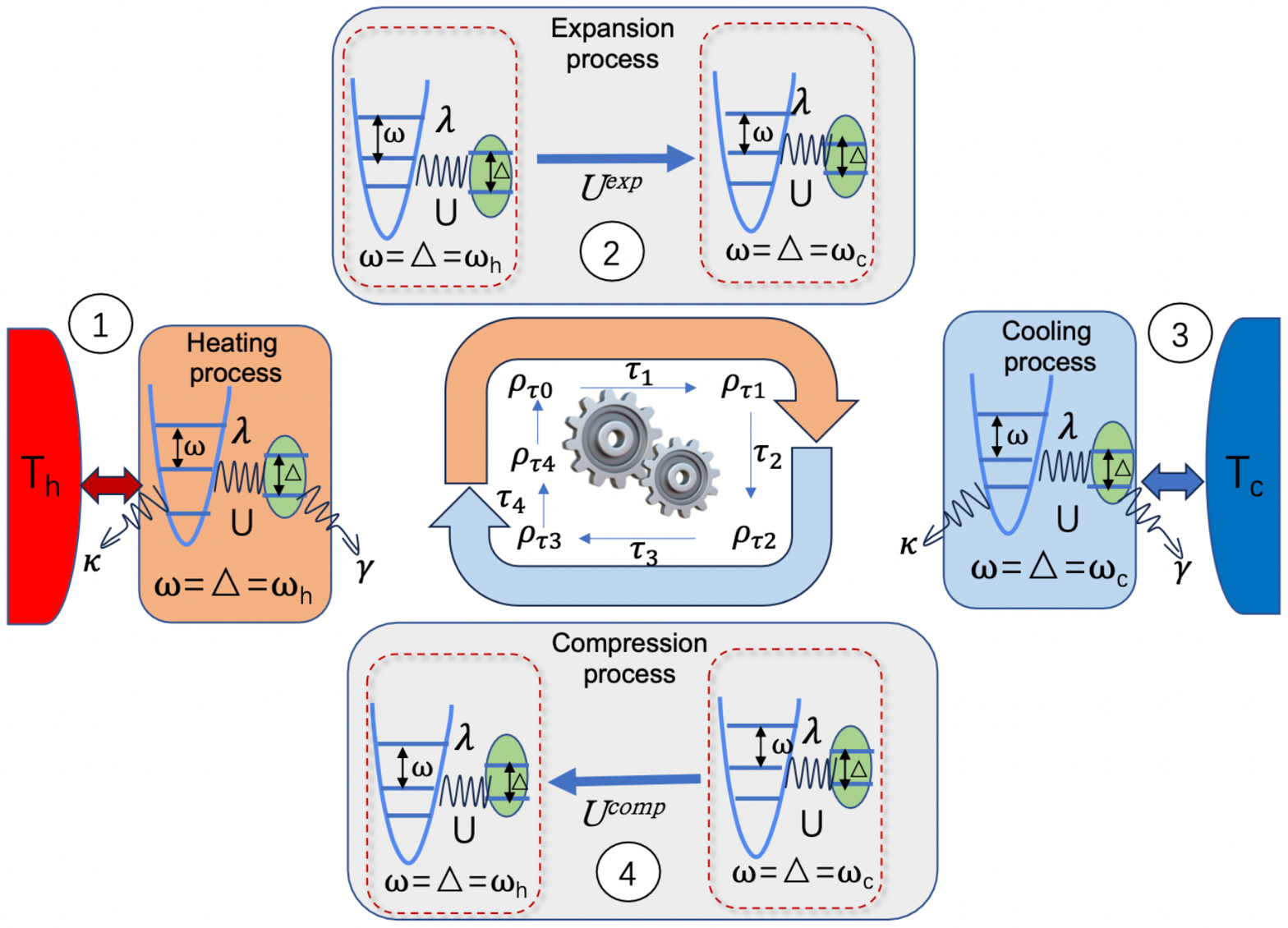}
%\vspace{-1.0cm}
\end{center}
\caption{Schematic representation of a Quantum Otto engine fueled by the anisotropic quantum Rabi-Stark model: \textbf{(1)} The working substance, described by Hamiltonian $H^{h}$ and frequency $\omega_{h}$, interacts with a hot reservoir at temperature $T_{h}$. \textbf{(2)} Isolated from the hot reservoir, energy levels change from $E_{n}^{h}$ to $E_{n}^{c}$ by adjusting the frequency from $\omega_{h}$ to $\omega_{c}$ ($\omega_{h} > \omega_{c}$). \textbf{(3)} With frequency $\omega_{c}$, the substance contacts a cold reservoir at temperature $T_{c} < T_{h}$. \textbf{(4)} Isolated from the cold reservoir, the system's energy levels revert from $E_{n}^{c}$ to $E_{n}^{h}$ by changing the frequency from $\omega_{c}$ to $\omega_{h}$, with work being performed.
}\label{Diagram1}
\end{figure}

A Quantum Otto engine operates by cyclically manipulating the work substance parameters to perform work, consisting of four main stages: compression, isochoric heating, expansion, and isochoric cooling\cite{quan2007quantum}. The thermodynamic cycle we consider is shown in
Fig.\ref{Diagram1} and proceeds as follows:

1. \textbf{Quantum Isochoric Heating Process}: 

The working substance, characterized by the Hamiltonian $H^{h}$ and frequency $\omega_{h}=\omega=\Delta$, comes into contact with a hot reservoir at temperature $T_{h}$. It's important to note that the AQRSM spectrum (see Fig.\ref{spectrum}) exhibits both non-degenerate and degenerate eigenvalues. These are adequately modeled by the secular approximation when deriving the Markovian master equation, except in cases of quasi-degeneracy. Following the approach outlined in Ref.\cite{mccauley2020accurate}, we can derive a thermodynamically valid master equation for the entire spectrum of the Hamiltonian, Eq.\eqref{eq:hamiltonian-aqrm}, up to the spectrum collapse at $|U|=1$. This derivation holds true under conditions of weak damping, a high bath cut-off frequency, and a flat spectral density, ensuring positivity and Markovianity. Under these conditions, the system evolves Markovianly according to the quantum master equation:

\begin{eqnarray}
   \frac{d}{dt}\hat{\rho} &=& -i[\hat{H}, \hat{\rho}] + \sum_{\substack{u = a, \sigma^{-} \\ k < j}} \left\{ \Gamma^{jk}_u n_u(\Delta_{jk}) \mathcal{D}[|\phi_j \rangle \langle \phi_k|, \hat{\rho}] \right. \nonumber \\
   && \left. + \Gamma^{jk}_u [1 + n_u(\Delta_{jk})] \mathcal{D}[|\phi_k \rangle \langle \phi_j|, \hat{\rho}] \right\},\label{eq:dressed-me}
   \end{eqnarray}
   where $\mathcal{D}[\hat{O}, \hat{\rho}] = \frac{1}{2} \left[ 2\hat{O}\hat{\rho}\hat{O}^{\dag} - \hat{\rho}\hat{O}^{\dag}\hat{O} - \hat{O}^{\dag}\hat{O}\hat{\rho} \right]$.
   
  The eigenvalues $E_k$ and eigenvectors $|\phi_k \rangle$ of the AQRSM Hamiltonian $ \hat{H} |\phi_k \rangle = E_k |\phi_k \rangle $ determine the system's dynamics. Dissipation rates $\Gamma^{jk}_u = \gamma_u(\Delta_{jk}) |S^{jk}_u|^2$ depend on the spectral function $\gamma_u(\Delta_{jk})$ And the transition coefficients: ${S}^{jk}_a = {\langle}\phi_j|(\hat{a}^{\dag}+\hat{a})|\phi_k{\rangle}$, ${S}^{jk}_{\sigma^{-}}={\langle}\phi_j|(\hat{\sigma}_++\hat{\sigma}_-)|\phi_k{\rangle}$. For the Ohmic case, $\gamma_u(\Delta_{jk}) = \pi \alpha \Delta_{jk} \exp(-|\Delta_{jk}| / \omega_{c})$, where $\alpha$ is the coupling strength and $\omega_{c}$ is the cutoff frequency, throughout all numerical simulations performed, we consider $\alpha=0.001\omega$ and $\omega_{c}=10\omega$. The Bose-Einstein distribution $n_u(\Delta_{jk}, T_u) = 1 / [\exp(\Delta_{jk} / T_u) - 1]$ accounts for thermal effects. The steady-state solution of Eq.\eqref{eq:dressed-me} yields the density matrix of the canonical ensemble, as confirmed by straightforward numerical simulations, given by:

\begin{eqnarray}
	\hat{\rho}_\mathrm{ss}=\sum_n\frac{e^{-E_n/T}}{\mathcal{Z}}|\phi_n{\rangle}{\langle}\phi_n|,\label{eq:ss}
\end{eqnarray}
where $\mathcal{Z}=\sum_ne^{-E_n/T}$ is the partition function, and the steady-state population is given by $P_{n}^{ss}=\frac{e^{-E_n/T}}{\mathcal{Z}}$.
At steady state, the population dynamics reach $\rho_{1} = \rho_{ss}(T_{h}) = \sum_{n} P_{n}^{ss}(T_{h}) |E_{n}^{h} \rangle \langle E_{n}^{h}|$, absorbing heat $Q_{h}$ without performing work.

2. \textbf{Quantum Adiabatic Expansion Process}: 
   The system, isolated from the hot reservoir, changes energy levels from $E_{n}^{h}$ to $E_{n}^{c}$ by varying the frequency from $\omega_{h}$ to $\omega_{c}$ ($\omega_{h} > \omega_{c}$). The populations $P_{n}^{ss}(T_{h})$ remain unchanged. The state becomes $\rho_{2} = \sum_{n} P_{n}^{ss}(T_{h}) |E_{n}^{c} \rangle \langle E_{n}^{c}|$. Only work is performed.

3. \textbf{Quantum Isochoric Cooling Process}: 
   The substance, with frequency $\omega_{c}$, contacts a cold reservoir at temperature $T_{c} < T_{h}$ until equilibrium is reached. The steady-state population changes to $P_{n}^{ss}(T_{c})$, and the state becomes $\rho_{3} = \sum_{n} P_{n}^{ss}(T_{c}) |E_{n}^{c} \rangle \langle E_{n}^{c}|$. Heat $Q_{c}$ is released, with no work done.

4. \textbf{Quantum Adiabatic Compression Process}: 
   Isolated from the cold reservoir, the system's energy levels revert from $E_{n}^{c}$ to $E_{n}^{h}$ by changing the frequency from $\omega_{c}$ to $\omega_{h}$. The state becomes $\rho_{4} = \sum_{n} P_{n}^{ss}(T_{c}) |E_{n}^{h} \rangle \langle E_{n}^{h}|$. Only work is performed.

%==========================================

\subsection{Work and Heat Calculations}

Using the first law of thermodynamics for a quantum system with discrete energy levels:
\begin{equation}
dU = \delta Q + \delta W = \sum_{n} (E_{n} dP_{n}^{ss} + P_{n}^{ss} dE_{n}),
\end{equation}
where $E_{n}$ are the energy levels and $P_{n}^{ss}$ are the steady-state occupation probabilities. The heat $Q_{h}$ and $Q_{c}$ exchanged with the hot and cold reservoirs, respectively, and the net work $W$ are given by:
\begin{eqnarray}
Q_{h} &=& \sum_{n} E_{n}^{h} [P_{n}^{ss}(T_{h}) - P_{n}^{ss}(T_{c})],
\label{Qhh}
\end{eqnarray}

\begin{eqnarray}
Q_{c} &=& \sum_{n} E_{n}^{c} [P_{n}^{ss}(T_{c}) - P_{n}^{ss}(T_{h})],
\label{Qc}
\end{eqnarray}
\begin{eqnarray}
W &=& Q_{h} + Q_{c} = \sum_{n} (E_{n}^{h} - E_{n}^{c}) [P_{n}^{ss}(T_{h}) - P_{n}^{ss}(T_{c})].
\label{W}
\end{eqnarray}

Positive values of $Q$ indicate heat absorption from a reservoir, while negative values indicate heat release. Similarly, positive values of $W$ indicate work done by the engine, and negative values indicate work done on the engine. The allowed working regimes under the Clausius inequality and the first law of thermodynamics are:

\begin{enumerate}
    \item \textbf{Heat engine (E)}: $Q_{h} > 0$, $Q_{c} < 0$, $W > 0$;
    \item \textbf{Refrigerator (R)}: $Q_{c} > 0$, $Q_{h} < 0$, $W < 0$;
    \item \textbf{Heater (H)}: $Q_{c} < 0$, $Q_{h} < 0$, $W < 0$;
    \item \textbf{Accelerator (A)}: $Q_{c} < 0$, $Q_{h} > 0$, $W < 0$.
\end{enumerate}

Among these four operation modes, the heat engine (E) and the refrigerator (R) are of the most practical interest. The heat engine operates by absorbing heat from a hot reservoir, converting part of this energy into work, and releasing the remaining heat to a cold reservoir. The efficiency of the heat engine, $\eta$, is a key figure of merit and is defined as the ratio of the work output to the heat absorbed from the hot reservoir $\eta = \frac{W}{Q_{h}}$.

In contrast, a refrigerator absorbs heat from a cold reservoir and releases it to a hot reservoir by consuming work. The performance of a refrigerator is measured by its coefficient of performance (COP), $\xi$, which is the ratio of the heat removed from the cold reservoir to the work input $\xi = \frac{Q_{c}}{|W|}.$

\subsection{Results Relevant to Our Analysis.}
 We review key findings from recent studies that provide a foundation for our discussion. These results play a crucial role in comprehending the mechanisms and behaviors observed in our work and help contextualize our findings within the broader field of quantum thermodynamics.
 To aid our analysis, it is useful to recall the analytical results for the quantum Otto cycle (QOC) when the working medium is a special multi-level system with all energy gaps changing by the same ratio in the quantum adiabatic process, i.e., $E_{n}^{h}-E_{m}^{h}=\alpha (E_{n}^{l}-E_{m}^{l})$, for $n=0,1,2,\ldots$ and $\alpha > 1$ since $E_{n}^{h} > E_{n}^{l}$. Substituting into the equations Eqs.(\ref{Qhh}) and (\ref{W}), we obtain the efficiency $\eta_{\alpha} = 1 - \frac{E_{n}^{l}-E_{m}^{l}}{E_{n}^{h}-E_{m}^{h}} = 1 - \frac{1}{\alpha}.$ In these instances, the quantum concept of adiabatic evolution aligns with the thermodynamic definition. This result, which stands for a special multi-level working fluid, generalizes the qubit engine and the quantum harmonic oscillator (QHO). For such systems, an adiabatic stroke maps an initial thermal state to another at a different temperature\cite{quan2007quantum}. 
In the limit of zero coupling for the AQRSM model, $\alpha=\omega_{h} /\omega_{c}$, the efficiency is given by $\eta_{\lambda =U= 0} = 1 - \omega_{c} /\omega_{h}$, and the coefficient of performance (COP) is $\xi_{\lambda =U= 0} = \omega_{c} /(\omega_{h}-\omega_{c})$, corresponding to a qubit and a non-interacting bosonic mode. Another important result is the positive-work condition (PWC) derived from the special multi-level harmonic spectrum given by $T_{h} > \frac{\omega_{h}}{\omega_{c}} T_{c}.$
Applying Eq.(\ref{W}) to the qubit engine yields $W_{\mathrm{qubit}} = (\omega_{h} - \omega_{c}) \left[ \frac{1}{1 + \exp(\beta_h \omega_{h})} - \frac{1}{1 + \exp(\beta_c \omega_{c})} \right]$, where $\beta_h = \frac{1}{k_B T_h}$ and $\beta_c = \frac{1}{k_B T_c}$ represent the inverse temperatures of the hot and cold reservoirs, respectively, with $k_B$ being the Boltzmann constant. For a bosonic mode, the total work done during the QOC is $W_{\mathrm{QHO}} = (\omega_{h} - \omega_{c}) \left[ \frac{1}{\exp(\beta_h \omega_h) - 1} - \frac{1}{\exp(\beta_c \omega_c) - 1} \right]$.

These expressions illustrate how the specific quantum nature of the working substance affects thermodynamic figures of merit, such as the engine's efficiency, by considering the respective energy gaps and temperature dependencies of the systems. However, in general, a quantum adiabatic stroke will drive a medium out of equilibrium, as occurs for AQRSM, although it will still remain in a passive energy-diagonal state.

The spectrum of the AQRSM displays a significant degree of anharmonicity and criticality, which are essential for understanding the factors influencing the efficiency and power of quantum Otto engines\cite{gelbwaser2018single}. This complexity makes it impractical to derive closed-form expressions for the thermodynamic quantities of the cycle, necessitating a numerical approach. Furthermore, examining the expressions for the decoupled system becomes crucial to assess the potential benefits arising from the interaction between radiation and matter in enhancing the thermal engine's performance. 

Criticality is an active area of exploration in both classical and quantum condensed matter physics, driven by the divergence of various parameters near critical points, which leads to universal behaviors in observed phenomena.
In a first-order phase transition, molar entropy and volume display discontinuity with temperature, while heat capacity diverges due to latent heat. Conversely, second-order phase transitions are characterized by a divergent susceptibility
\cite{chaikin1995principles}.
It has been theorized that near a classical phase transition, the divergence of equilibrium fluctuations in a second-order QPT, as indicated by the fluctuation-dissipation theorem, may enhance heat engine performance\cite{campisi2016power}. At zero temperature, a system can undergo a quantum phase transition, leading to changes in non-analyticities in its ground state.
Evidence supporting this quantum critical enhancement in engine operation has been found in several systems undergoing continuous phase transitions\cite{Fogarty_2021, 10.1088/1367-2630/ac963b, PhysRevE.96.022143, interaction_chen, PhysRevResearch.2.043247, e24101458, free_evolution, xu2024universal, mukherjee2024quantum}. Additionally, such singular behaviors may be reflected in stochastic or quantum thermodynamic quantities, as the singular behavior of entropy production indicates a second-order transition\cite{mascarenhas2014work, dorner2012emergent, zawadzki2020work}. However, the implications of first-order QPT in heat engines are not fully understood.

First-order quantum phase transitions are intriguing due to potential differences in the characteristic relaxation frequency spectrum of fluctuations and alterations in the nature of quantum coherence near the transition point\cite{ pfleiderer2005first}.
These aspects imply that the realm of first-order quantum phase transitions offers ample opportunities for both theoretical exploration and experimental investigation in the domain of quantum thermodynamics. It's noteworthy to mention that in a quenched system experiencing a first-order QPT, the average work exhibits discontinuity\cite{ mascarenhas2014work}.

Finally, we should mention that the excited-state quantum phase transition (ESQPT) was thoroughly studied for the QRM \cite{puebla2016excited} and AQRM \cite{hu2023excited} in the limit of infinite frequency ratios ($\Delta/\omega \to \infty$). Excited-state quantum phase transitions (ESQPTs) refer to non-analytic changes in the properties of excited states of a quantum system as a control parameter is varied. Unlike ground-state quantum phase transitions, which occur at zero temperature, ESQPTs manifest at finite energy densities and are characterized by singularities in the density of states. These transitions have significant implications for thermodynamic quantities such as entropy, specific heat, and work. For instance, near an ESQPT, the system's specific heat may exhibit anomalous peaks, and the response functions can show critical scaling behavior\cite{cejnar2006monodromy,Relano2008}. This criticality can influence the efficiency and performance of quantum heat engines, as the presence of ESQPTs may enhance the engine's work output and alter the cycle's thermodynamic properties. Understanding ESQPTs thus provides deeper insights into the non-equilibrium dynamics and the thermodynamic behavior of quantum systems, potentially leading to optimized designs for quantum thermal engines\cite{Cejnar2021}. However, to the best of our knowledge, there have been no studies on ESQPTs in the AQRSM at finite frequency ratios. Investigating ESQPTs in the AQRSM could provide valuable insights into the behavior of quantum engines operating under these conditions, potentially revealing new avenues for enhancing engine performance and efficiency in practical applications.

\subsection{Operation in Finite Time}

In our investigations thus far, we have primarily focused on analyzing the engine's performance in the adiabatic limit, where the power output tends towards zero due to the extended duration of the strokes. Although reducing the duration of the strokes can increase power, it inevitably introduces inefficiencies due to the generation of irreversibility and internal friction\cite{e19040136,shiraishi2016universal}.
The emergence of quantum friction presents a significant challenge in finite-time quantum processes. Quantum friction arises from non-adiabatic excitations during rapid driving of the working substance, leading to energy dissipation and reduced efficiency. Ongoing efforts are directed towards comprehending and mitigating quantum friction, with the aim of enhancing the performance of quantum heat engines operating within finite time frames\cite{rezek2006irreversible,Fogarty_2021}.

To comprehend the trade-off between power and efficiency in the engine, our attention is directed towards the heat engine \textbf{E}, which begins with the system in a thermal state at temperature $T_h$.
Let us consider a dynamic system with a generic time-dependent Hamiltonian $H(t)$. The average energy, $U(t)$, for state $\rho(t)$ is $U(t) = \langle H(t) \rangle = \text{Tr}\{\rho(t) H(t)\}$, with the rate of change of system energy  $\frac{dU}{dt} = \frac{dW}{dt} + \frac{dQ}{dt}$. Here, $Q$ and $W$ represent heat and work, respectively, and are given by:
\begin{equation}
Q(t) = \int_{0}^{t} dt' \text{Tr}\{H(t') \dot{\rho}(t')\},
\label{heatdef}
\end{equation}
\begin{equation}
W(t) = \int_{0}^{t} dt' \text{Tr}\{\dot{H}(t') \rho(t')\}.
\label{workdef}
\end{equation}
For the finite time quantum Otto engine cycle process, we 
 set the working substance begins in the hot Gibbs state $\rho_{\tau_{0}}$. During an isochoric process, see Fig.\ref{Diagram1} strokes $\mathbf{1}$ and $\mathbf{3}$, where $ H(t) = H $ and $ W(t) = 0 $, both thermalization times are set equal, denoted as $ \tau_2 = \tau_4 $. Adiabatic processes, see Fig.\ref{Diagram1} strokes $\mathbf{2}$ and $\mathbf{4}$, where $ Q(t) = 0 $, are implemented by linearly varying the working substance frequency as $ \omega(t) = \omega_j - \tau_j t $, with $ t $ ranging from $ 0 $ to $ \frac{\omega_f - \omega_i}{\tau_n} $, where $ j = h, c $ for $ n = 1, 3 $, and $ \tau_1 = \tau_3 $. For the ideal quantum Otto engine cycle, $\tau_1=\tau_3\rightarrow\infty$, $\tau_2=\tau_4\rightarrow\infty$, and  $\rho_{\tau_{4}}=\rho_{\tau_{0}}$.
Throughout the four stages of the quantum Otto cycle, energy exchange occurs solely through work or heat. The total extracted work $ W(t) $ is the sum of $ Q_{1}(t) $ and $ Q_{3}(t) $, yielding the efficiency $ \eta(t) = \frac{W(t)}{Q(t)} $. The power $ P(t) $ is the derivative of work with respect to time, calculated as $ \frac{dW}{dt} = \text{Tr}\{\dot{H}(t) \rho(t)\} $.

\begin{figure}[tbp]
\begin{center}
%\vspace{-2.2cm}
\includegraphics[width=0.4\textwidth]{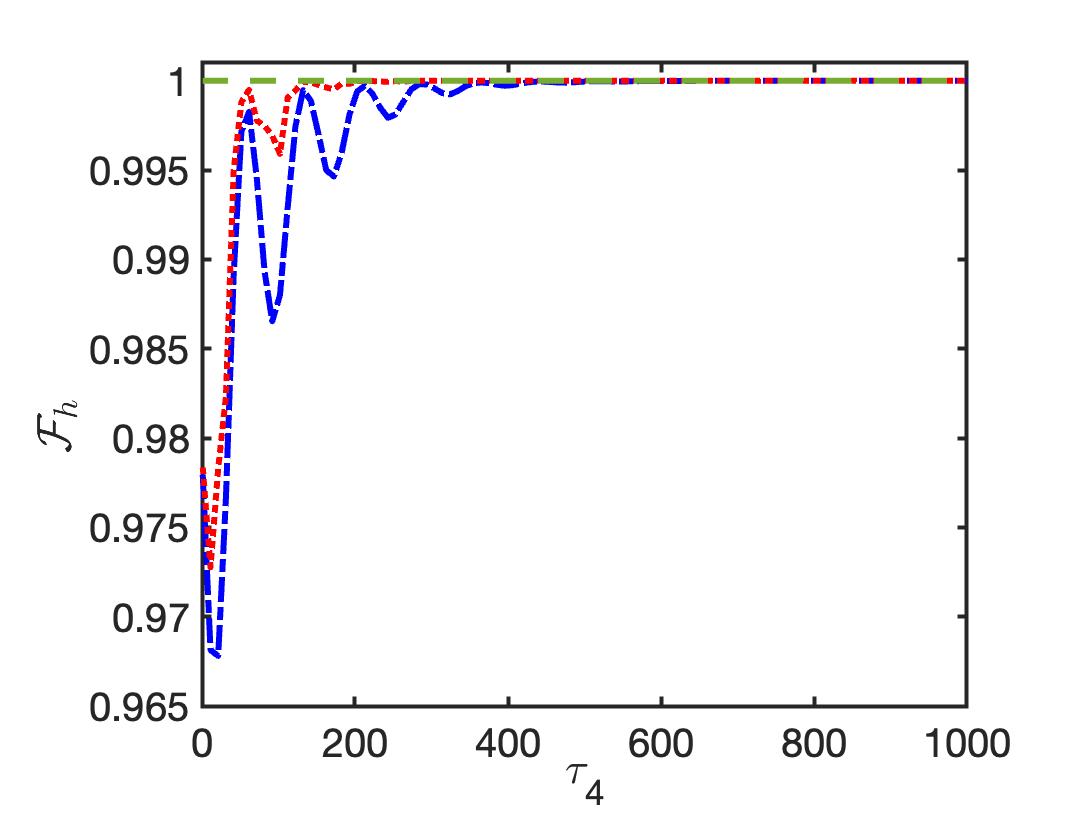}
%\vspace{-1.0cm}
\end{center}
\caption{Fidelity betweem $\rho^{(N)}(\tau_0)$ and $\rho^{(N-1)}(\tau_0)$ as a function of the cycle time $\tau_{4}$ where $N=2$ (dash dot  blue), $N=3$ (red dotted line), $N=4$ (green dashed line). 
The other system parameters are given by $T_{h}=0.5$, $T_{c}=0.1$, $\lambda_1=\lambda_2=0.5$, $U=0$. All quantities above are in units of $\omega$.
}\label{Fidelity}
\end{figure}

We investigate the quantum heat engine (QHE) through successive iterations of the cycle. The isochoric stages entail Markovian evolution, as illustrated in Fig.\ref{Fidelity}, we observe that the system converges to a limit cycle irrespective of the initial state. We begin each cycle with the hot Gibbs state $\rho_{\tau_{0}}$ as our initial condition. Upon completing the cycle, the system settles into the final state $\rho_{\tau_{4}}$. This final state then serves as the starting point $\rho_{\tau_{0}}$ for the subsequent cycle, continuing this iterative process. As a result, multiple iterations of the cycle are performed.
In Fig. \ref{Fidelity}, we illustrate the fidelity $F_{h} = F(\rho^{(N-1)}(\tau_{0}), \rho^{(N)}(\tau_{0}))$, which quantifies the similarity between the state $\rho(\tau_{0})$ across consecutive iterations. The fidelity $F$ equals 1 when the states are identical. Notably, for each cycle, when $\tau_{4} > 800$, the fidelity $F_{h} \rightarrow 1$. After numerous iterations, the system's state exhibits minimal variation during the cycle, with $F_{h} \rightarrow 1$. This observation indicates that during very short cycle times, the system's state remains nearly unchanged (see Fig. \ref{Fidelity}).

For short time of $\tau_4$, the evolution through one full cycle changes the state of the system, leading to a decrease in the fidelity between consecutive cycles. This decrease indicates that the system's state is evolving and not remaining in a steady state during these short but non-negligible cycle times.

As we consider increasing values of $\tau_{4}$, the interaction with the hot reservoir drives $\rho(\tau_{4})$ closer to the stationary thermal state. This thermalization process increases the fidelity because the system spends more time in contact with the reservoir, allowing it to reach equilibrium. For times comparable with the relaxation time, $F_{h} \rightarrow 1$, indicating that the engine operates in the limit cycle from the first iteration. This behavior is expected, as longer interaction times with the reservoir allow the system to reach and maintain thermal equilibrium more efficiently.

Interestingly, the limit cycle can also be achieved with a nonequilibrium steady state. In such cases, the system reaches a steady-state that is not in thermal equilibrium but remains constant over time under continuous operation. This steady state can be maintained by balancing the energy exchanges with the reservoirs, ensuring the system's stability and consistent performance. 

Throughout this work, we characterize the thermodynamic figures of merit in the limit cycle as defined above. The limit cycle behavior is crucial for understanding the steady-state performance of the quantum Otto engine, ensuring consistent and reliable operation.

To assess the detrimental effects of finite-time operation, we consider the total entropy produced in a cycle, denoted as $\left\langle \Sigma_{\text{total}} \right\rangle$. This measure encompasses both the finite-time effects of the engine's dynamics and the thermalization process, directly reflecting the irreversibility within the cycle. The total entropy production can be expressed as
\begin{equation}
\begin{split}
    \langle\Sigma_{\text{total}}\rangle & = -\beta_{h}[\text{Tr}(\rho_{\tau_{4}}H_{h})-\text{Tr}(\rho_{\tau_{3}}H_{h})]\\
    &  -\beta_{c}[\text{Tr}(\rho_{\tau_{2}}H_{c})-\text{Tr}(\rho_{\tau_{1}}H_{c})].\label{eq:entropy}
\end{split}
\end{equation}
As a complementary measure of the detrimental effects of finite-time operations, we quantify the work due to internal friction using the Kullback-Leibler divergence
\begin{equation}
W_{\text{fric}} = \frac{1}{\beta_{h(c)}} D(\rho_{1(3)} || \rho_{1(3)}^{\text{qe}}),\label{eq:work}
\end{equation}
where $D(\rho || \rho^{\text{ref}}) = \text{Tr}[\rho (\ln \rho - \ln \rho^{\text{ref}})]$. Here, $\rho_{1(3)}^{\text{qe}}$ denotes the quasistatically evolved state corresponding to $\rho_{1(3)}$. The total friction work sums contributions from compression and expansion strokes $W^{\text{fric}} = W_{\text{comp}}^{\text{fric}} + W_{\text{exp}}^{\text{fric}}$.

%The entropy produced during incomplete thermalization with a Markovian reservoir is:

%\begin{equation}
%\left\langle \Sigma \right\rangle = D(\rho_0 || \rho^{\text{eq}}) - D(\rho_{\tau} || \rho^{\text{eq}}) \geq 0,
%\end{equation}
%where $\rho_0$ and $\rho_{\tau}$ are initial and final states of the %thermalization process, respectively, and $\rho^{\text{eq}}$ is the %Gibbs state.

An important result that warrants attention is that the efficiency of the quantum engine can be related to the total entropy produced in a cycle as \cite{PhysRevLett.123.240601}

\begin{equation}
\eta = \eta_{\text{Carnot}} - \frac{\left\langle \Sigma_{\text{total}} \right\rangle}{\beta_{\text{c}} \left\langle Q_{\text{h}} \right\rangle},
\end{equation}
where $ \eta_{\text{Carnot}} = 1 - \beta_{\text{h}} / \beta_{\text{c}} $ is the Carnot efficiency. The quantity $ \mathcal{L}_{\text{therm}} = \left\langle \Sigma_{\text{total}} \right\rangle / \beta_{\text{c}} \left\langle Q_{\text{h}} \right\rangle $ quantifies the departure of the engine's efficiency from the Carnot efficiency. Since $ \left\langle \Sigma_{\text{total}} \right\rangle $ is nonnegative, the quantum engine's efficiency is always upper-bounded by the Carnot efficiency. 

Furthermore, total entropy  production thisthis$\langle\Sigma_{\text{total}}\rangle$ plays a crucial role in quantifying the relative fluctuations of integrated currents, such as heat or energy, through thermodynamic uncertainty relations\cite{pietzonka2018universal,falasco2020unifying,timpanaro2019thermodynamic}. These relations establish a trade-off between optimizing relative fluctuations and entropy production, providing constraints that are universally applicable across different scenarios. Consequently, a continuous heat engine operating within a non-equilibrium steady state must adhere to a trade-off relationship involving its efficiency, output power, and power fluctuations. This relationship can be expressed through the integrated current $\mathcal{Q}_i$, where $\mathcal{Q}_i$ represents quantities like work ($W$), efficiency ($\eta$), or power ($P$)\cite{timpanaro2019thermodynamic}:

\begin{equation}\label{Timpa}
\frac{\var(\mathcal{Q}_i)}{\langle \mathcal{Q}_i \rangle^2} \geq f(\langle \Sigma_{total} \rangle),
\end{equation}
where $f(x) = \text{csch}^2(g(x/2))$. Here, $\text{csch}(x)$ represents the hyperbolic cosecant function, and $g(x)$ is the inverse function of $x \tanh(x)$.

%==========================================
%=================

\begin{figure}[!htbp]
\begin{center}
%\vspace{-2.2cm}

\includegraphics[width=0.5\textwidth]{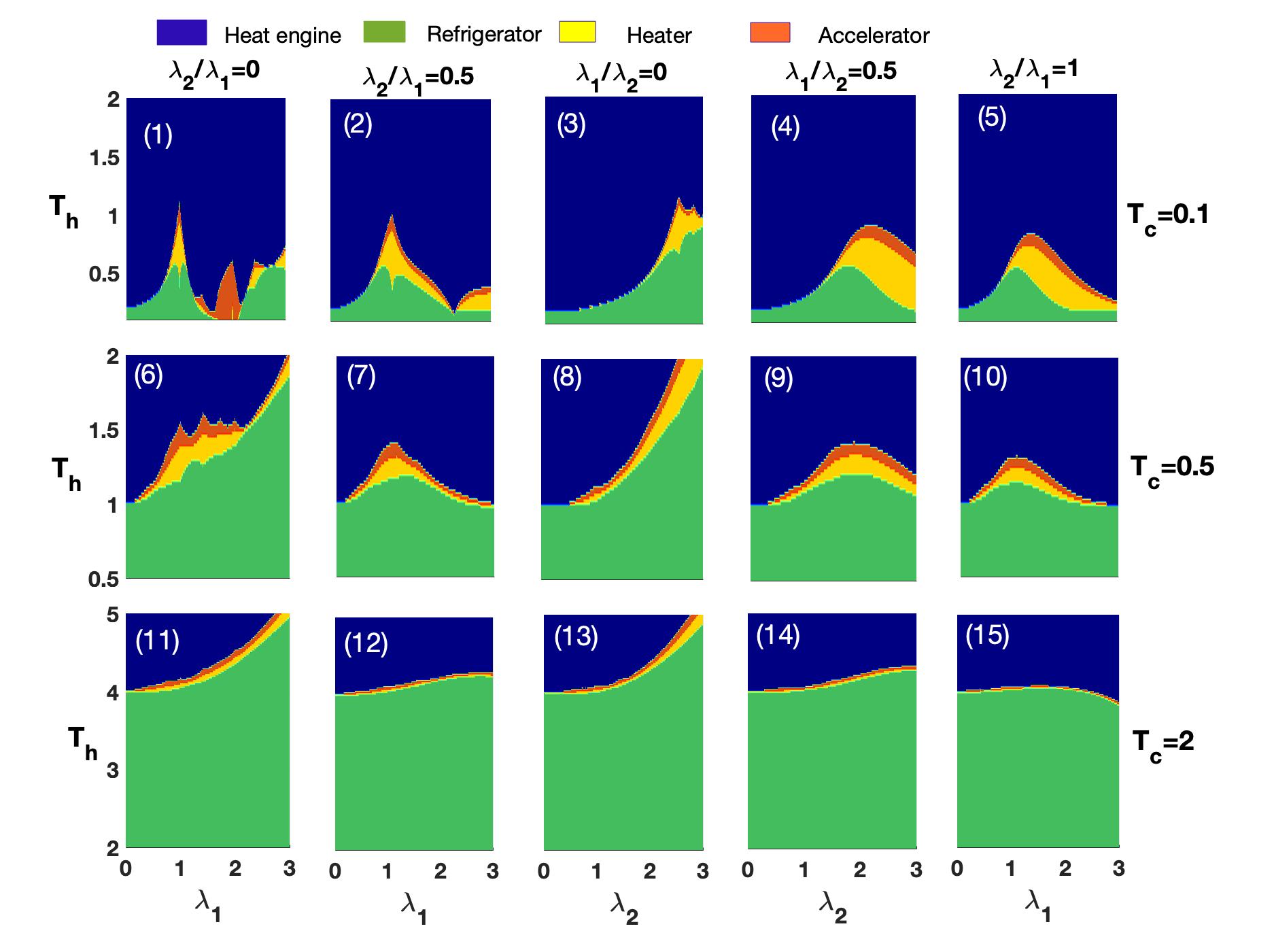}
%\vspace{-1.0cm}
\end{center}
\caption{
Phase diagram illustrating the operating regimes of the quantum Otto machine. The diagram shows how the hot reservoir temperature ($T_h$) and the qubit-boson coupling strengths ($\lambda_1, \lambda_2$) influence the machine's behavior, with a fixed ratio of $\lambda_2/\lambda_1$ (or $\lambda_1/\lambda_2$). The cold reservoir temperatures are set at $T_c = 0.1$ (panels 1-5), $T_c = 0.5$ (panels 6-10), and $T_c = 2$ (panels 11-15). The color coding represents different operating modes: heat engine (deep blue), refrigerator (green), heater (yellow), and accelerator (red). All quantities are in units of $\omega$, with parameters $U = 0$, $\omega_h = 2\omega$, and $\omega_c = \omega$.}\label{3DThTc}
\end{figure}
%==========================================
\begin{figure}[!htbp]
\centering
%\vspace{-2.2cm}
\includegraphics[width=0.5\textwidth]{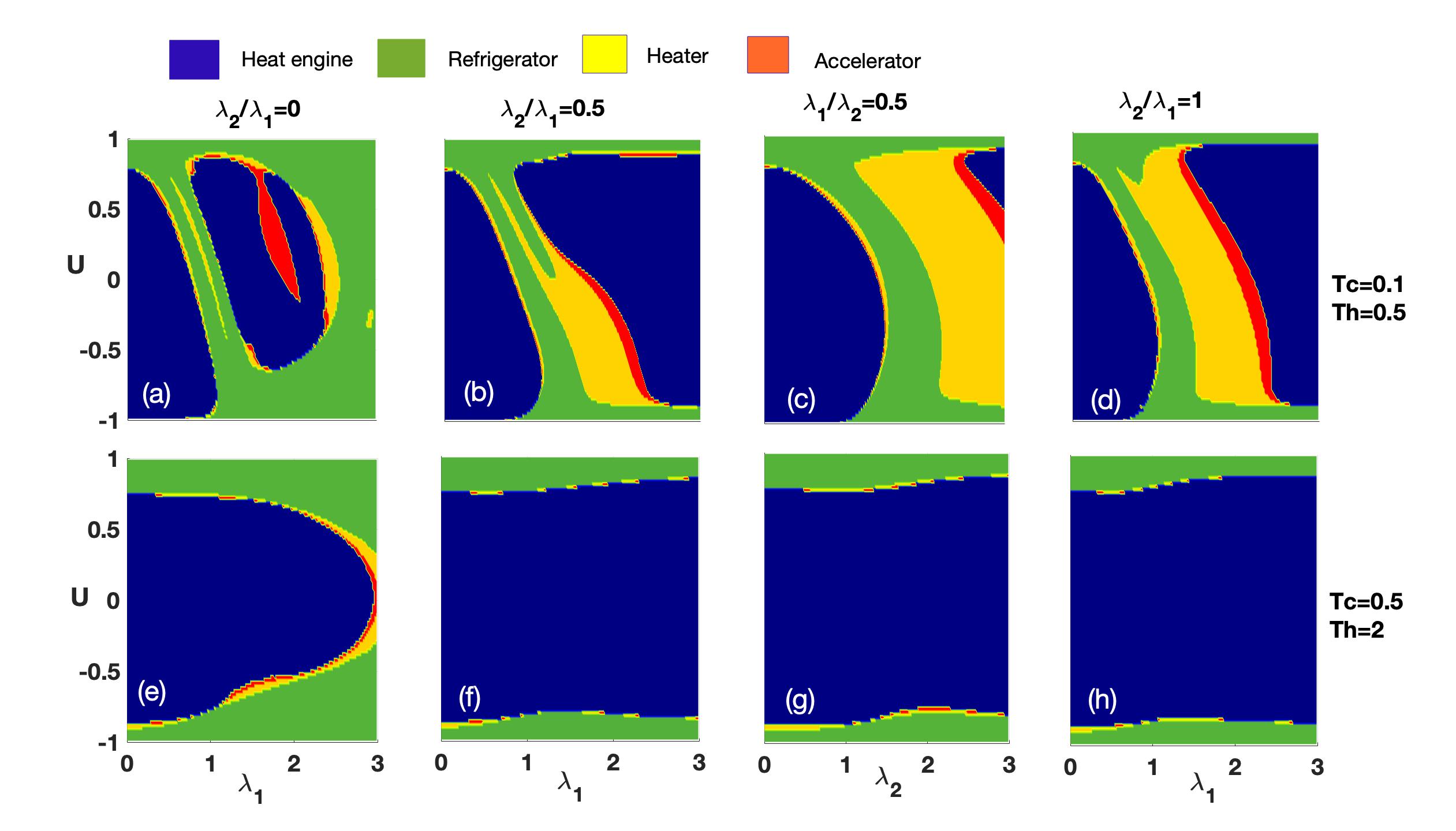}
%\vspace{-1.0cm}

\caption{Operating regimes of the quantum Otto machine with varying nonlinear Stark coupling strength $U$ and qubit-boson coupling strengths $\lambda_{1} (\lambda_{2})$, both in units of $\omega$. Fixed coupling ratios are: (a),(e) $\lambda_{2}/\lambda_{1}=0$, (b),(f) $\lambda_{2}/\lambda_{1}=0.5$, (c),(g) $\lambda_{1}/\lambda_{2}=0.5$, and (d),(h) $\lambda_{1}/\lambda_{2}=1$. Color codes indicate different operational modes: heat engine (deep blue), refrigerator (green), heater (red), and accelerator (yellow). Parameters: $T_{h}=0.5$, $T_{c}=0.1$ (first row); $T_{h}=2$, $T_{c}=0.5$ (second row); $\omega_{h}=2\omega$, $\omega_{c}=\omega$.
}\label{Ulambda}
\end{figure}

\begin{figure*}[htbp]
    \centering
    \begin{subfigure}[b]{0.45\textwidth}
        \centering
        \includegraphics[width=\textwidth]{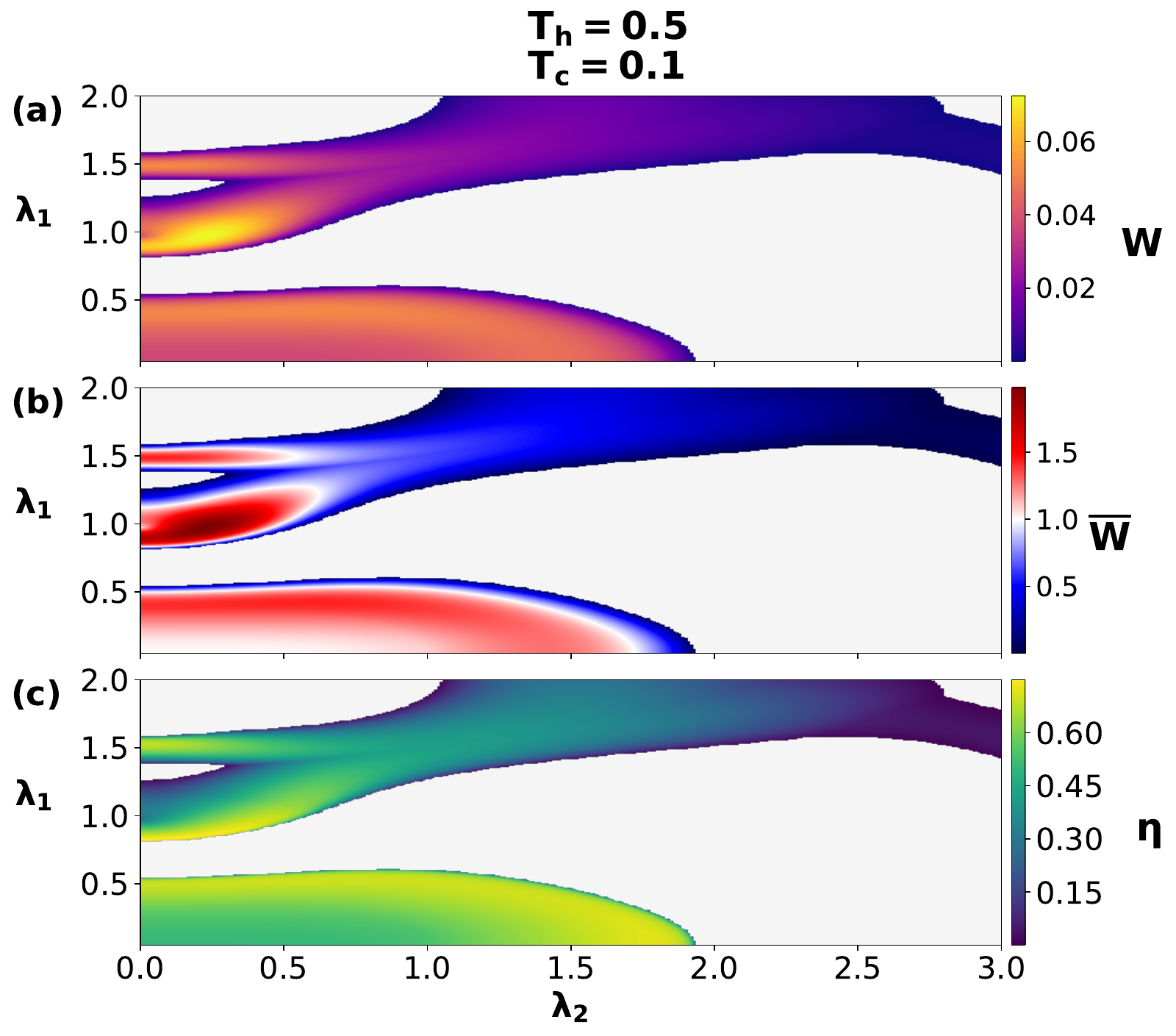}
        \end{subfigure}
    \hfill
    \begin{subfigure}[b]{0.45\textwidth}
        \centering
        \includegraphics[width=\textwidth]{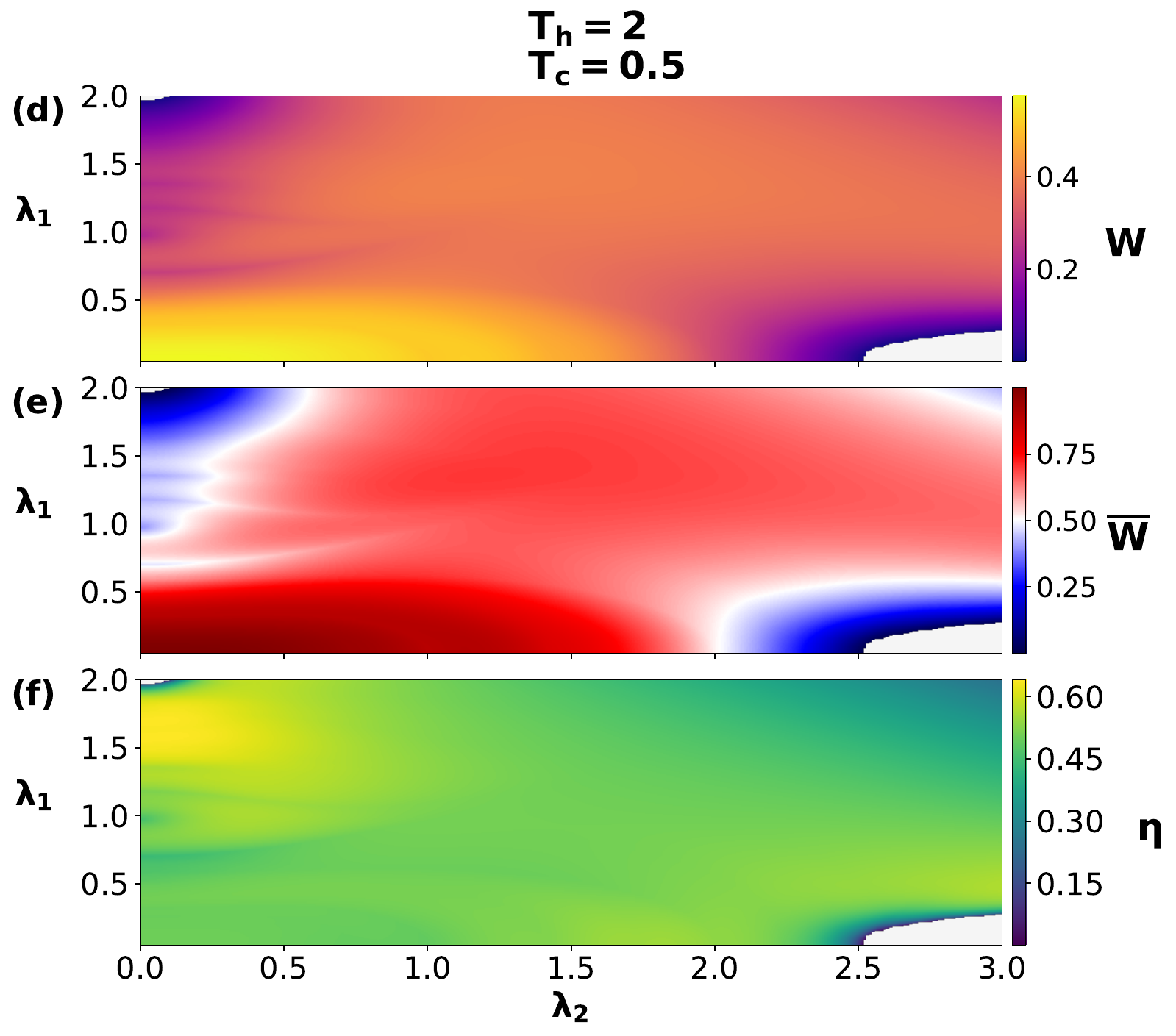}
            \end{subfigure}
    \caption{2D heatmaps depicting the Work (W), the normalized work $\overline{W}$, and  Efficiency (\(\eta\)) of the Quantum Otto Cycle powered by the anisotropic quantum Rabi model. Panels (a) and (d) show the net Work and Panels (b) and (e) show the  normalized work, while panels (c) and (f) display the Efficiency as functions of the coupling strengths \(\lambda_1\) (RW) and \(\lambda_2\) (CRW). The left column presents results for a cold reservoir temperature \(T_c = 0.1\) and a hot reservoir temperature \(T_h = 0.5\). The right column shows results for \(T_c = 0.5\) and \(T_h = 2\). The color bars indicate the magnitude of the Work and Efficiency. Cycle parameters include \(\omega_h = 2\omega\) and \(\omega_c = \omega\), with all quantities measured in units of \(\omega\). Gray areas represent regimes where the system does not operate as a heat engine.
}
    \label{2DAQRM}
\end{figure*}

\begin{figure*}[htbp]
    \centering
    \begin{subfigure}[b]{0.45\textwidth}
        \centering
        \includegraphics[width=\textwidth]{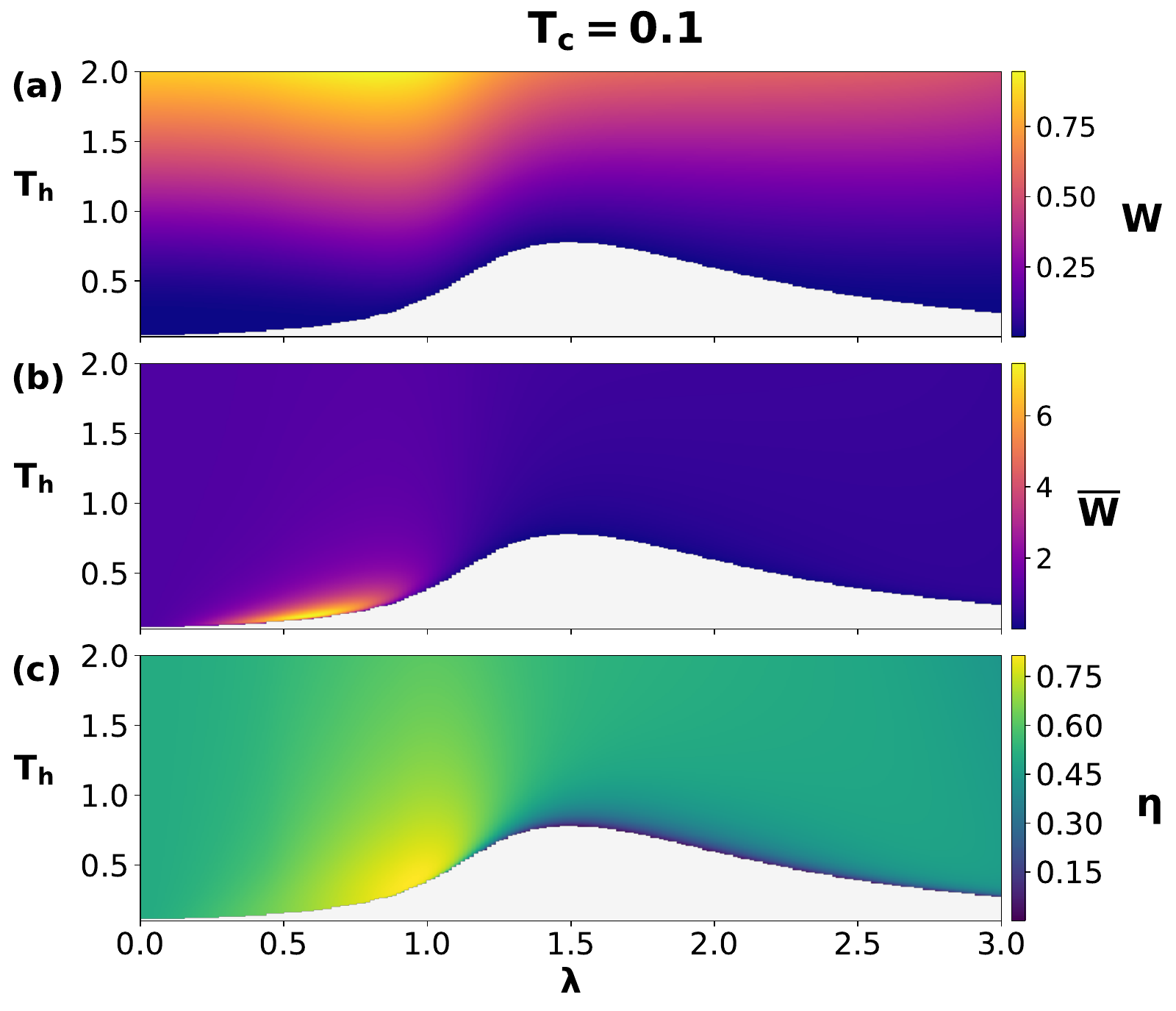}
        \end{subfigure}
    \hfill
    \begin{subfigure}[b]{0.45\textwidth}
        \centering
        \includegraphics[width=\textwidth]{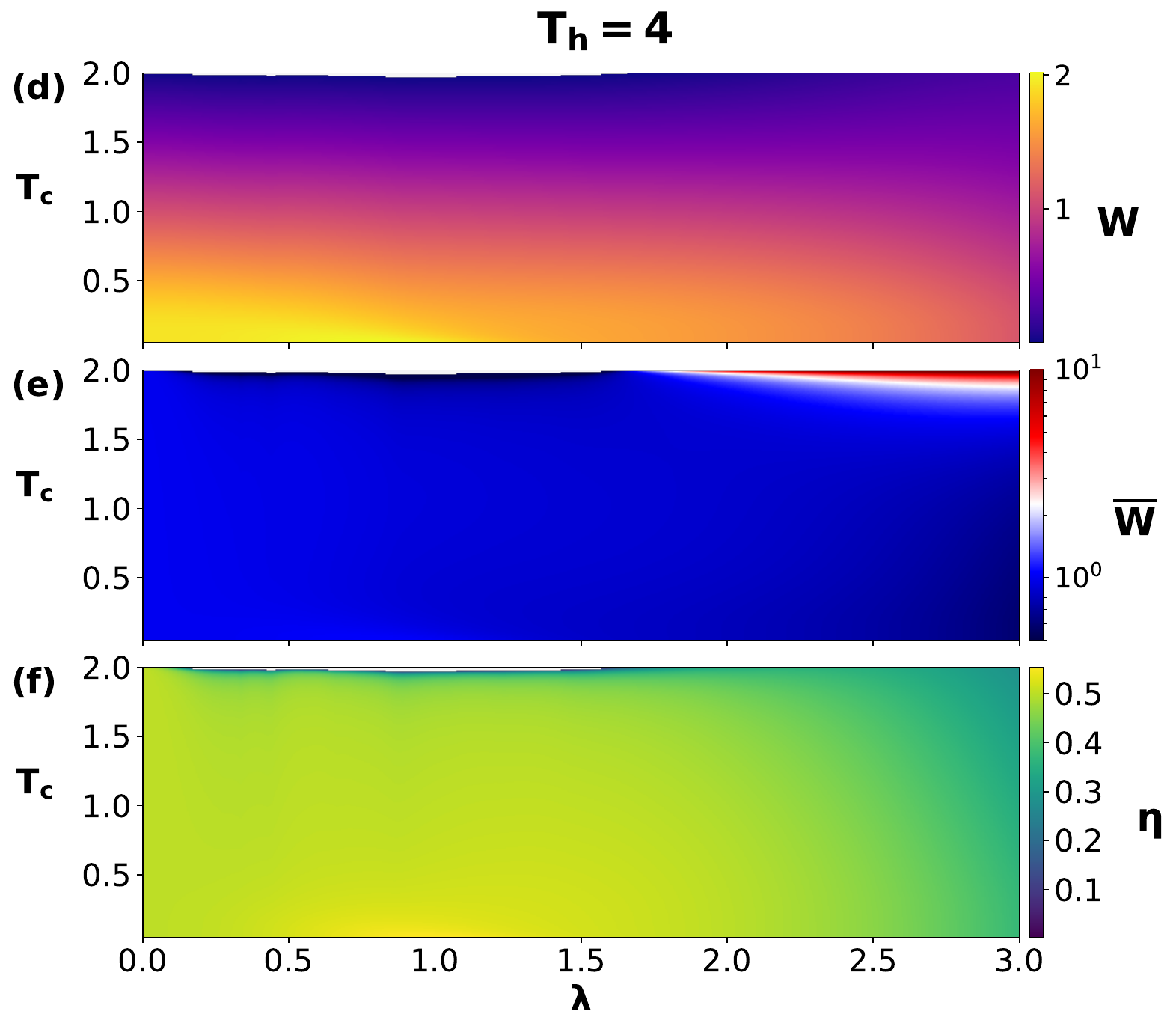}
            \end{subfigure}
    \caption{2D heatmaps depicting the Work (W), the normalized work $\overline{W}$, and  Efficiency (\(\eta\)) of the Quantum Otto cycle powered by the quantum Rabi model. The heatmaps show the dependence of Work (a, d), the normalized work (b, e) and Efficiency (c, f) on the coupling strength $\lambda_1 = \lambda_2 = \lambda$. The left column presents the results for a cold reservoir with fixed temperature \(T_c = 0.1\). The right column shows the results for a hot reservoir with fixed temperature \(T_h = 4\). Color bars indicate the magnitude of the work, normalized work, and Efficiency. The cycle parameters include \(\omega_h = 2\omega\) and \(\omega_c = \omega\), with all quantities measured in units of \(\omega\). Gray areas represent regimes where the system does not operate as a heat engine.}

    \label{2DAQRM2}
\end{figure*}

\begin{figure*}[htbp]
    \centering
    \begin{subfigure}[b]{0.45\textwidth}
        \centering
        \includegraphics[width=\textwidth]{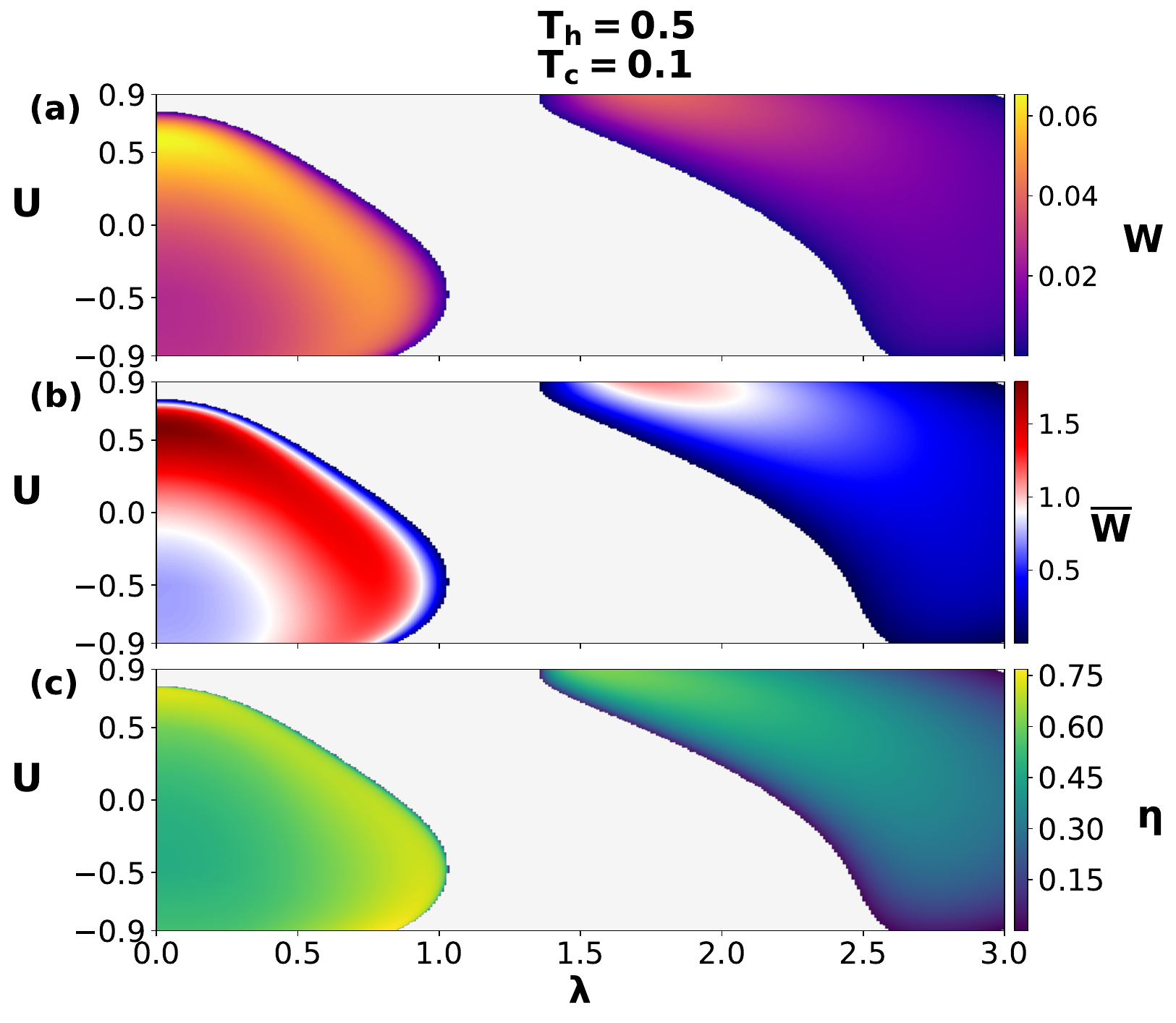}
        \end{subfigure}
    \hfill
    \begin{subfigure}[b]{0.45\textwidth}
        \centering
        \includegraphics[width=\textwidth]{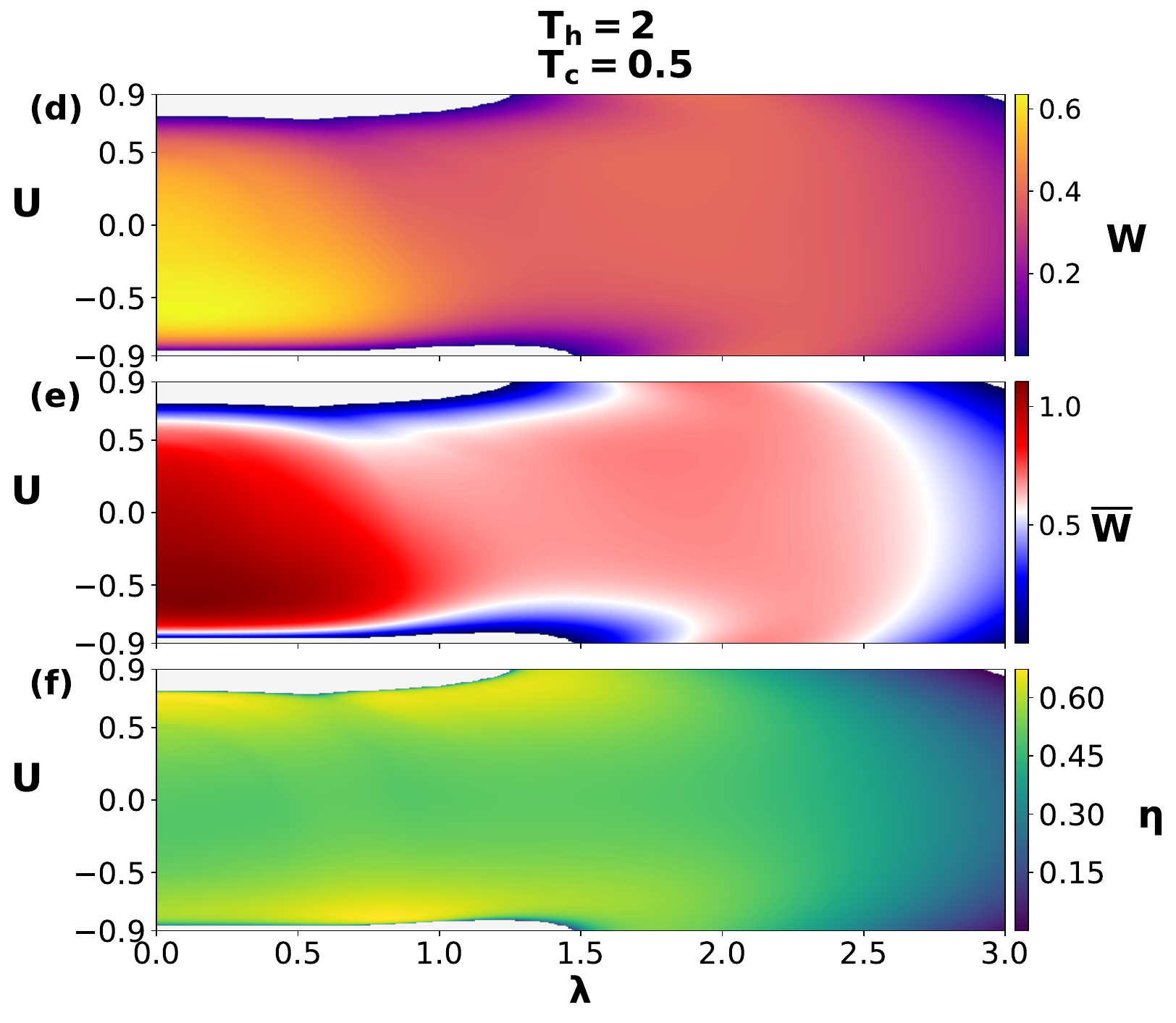}
            \end{subfigure}
    \caption{2D heatmaps depicting the Work W (a,d), the normalized work $\overline{W}$ (b,e), and  Efficiency (\(\eta\)) (c,f) as functions of the Stark interaction $U$ and the coupling $\mathbf{\lambda}$ for a Quantum Otto Cycle powered by the quantum Rabi-Stark model. The left column presents the results for cold and hot temperatures $ T_c = 0.1 $ and $ T_h = 0.5 $, while the right column displays the results for $ T_c = 0.5 $ and $ T_h = 2 $. Color bars indicate the magnitude of the work, normalized work, and Efficiency. For reference, the qubit and quantum harmonic oscillator engine achieve a constant efficiency of $\eta = 0.5$, with work $W_{\mathrm{qubit}} \approx 0.018(0.15)$ and $W_{\mathrm{QHO}} \approx 0.019(0.43)$, respectively, for the parameters in the first (second) column. Other cycle parameters are $\omega_{h} = 2\omega$ and $\omega_{c} = \omega$, with all quantities in units of $\omega$. Gray areas represent regimes where the system does not operate as a heat engine.}
    \label{combined_figures}
\end{figure*}

%==========================================

%==========================================
\begin{figure*}[tbp]
\begin{center}
%\vspace{-2.2cm}
\includegraphics[width=0.9\textwidth]{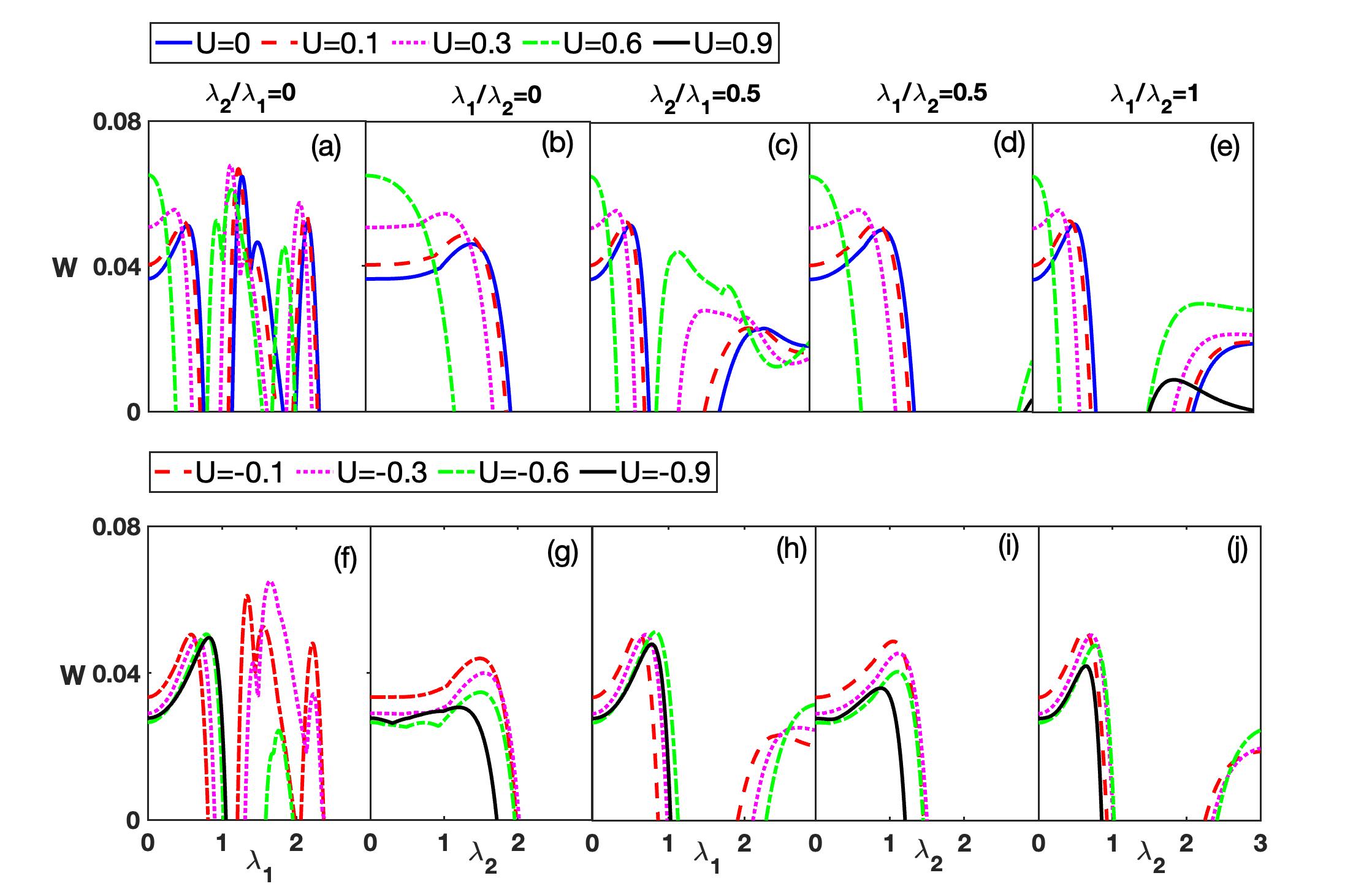}
%\vspace{-1.0cm}
\end{center}
\caption{(a)-(e) Work W extracted from the quantum heat engine as a function of the qubit-boson coupling strength ratio $\lambda_{1}/\lambda_{2}$ for different nonlinear Stark coupling strength $U=0$ (solid blue line), $U=0.1$ (dashed red line), $U=0.3$ (dotted magenta line), $U=0.6$ (dash dot  green line), and $U=0.9$ (solid black line) with fixed coupling ratio (a)$\lambda_{2}/\lambda_{1}=0$,(b)$\lambda_{1}/\lambda_{2}=0$,(c)$\lambda_{2}/\lambda_{1}=0.5$, (d)$\lambda_{1}/\lambda_{2}=0.5$ and (e)$\lambda_{2}/\lambda_{1}=1$. (f)-(j) Work extracted from the quantum heat engine as a function of the qubit-boson coupling strength $\lambda_{1}$($\lambda_{2}$) in units of $\omega$ for different nonlinear Stark coupling strength $U=-0.1$ (dashed red line), $U=-0.3$ (dotted magenta line), $U=-0.6$ (dash dot  green line), $U=-0.9$ (solid black line) with fixed coupling ratio (f) $\lambda_{2}/\lambda_{1}=0$,(g) $\lambda_{1}/\lambda_{2}=0$,(h) $\lambda_{2}/\lambda_{1}=0.5$, (i) $\lambda_{1}/\lambda_{2}=0.5$ and (j )$\lambda_{2}/\lambda_{1}=1$. 
The other system parameters are given by $T_{h}=0.5$, $T_{c}=0.1$,
$\omega_{h} = 2\omega$, $\omega_{c} = \omega$. All quantities above are in units of $\omega$.
}\label{Stark9}
\end{figure*}

\begin{figure*}[tbp]
\begin{center}
%\vspace{-2.2cm}
\includegraphics[width=0.9\textwidth]{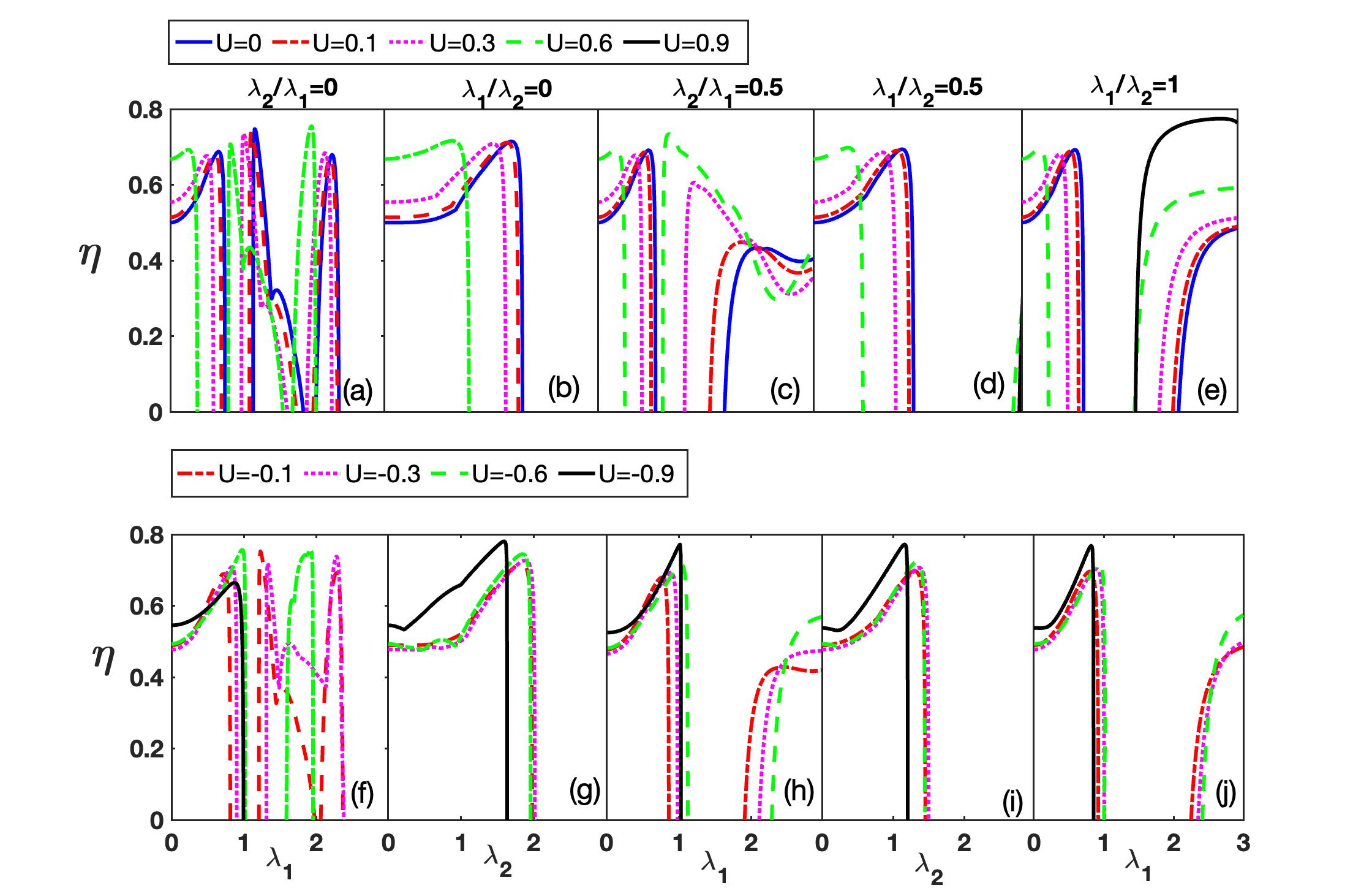}
%\vspace{-1.0cm}
\end{center}
\caption{Efficiency $\eta$ of the quantum heat engine as a function of the qubit-boson coupling strength ratio $\lambda_{1}/\lambda_{2}$ for nonlinear Stark coupling strengths: (a)-(e) $U=0$ (solid blue line), $U=0.1$ (dashed red line), $U=0.3$ (dotted magenta line), $U=0.6$ (dash dot  green line), and $U=0.9$ (solid black line) fixing (a) $\lambda_{2}/\lambda_{1}=0$,(b) $\lambda_{1}/\lambda_{2}=0$,(c) $\lambda_{2}/\lambda_{1}=0.5$, (d) $\lambda_{1}/\lambda_{2}=0.5$ and (e) $\lambda_{2}/\lambda_{1}=1$;  (f)-(j) $U=-0.1$ (dashed red line), $U=-0.3$ (dotted magenta line), $U=-0.6$ (dash dot  green line), $U=-0.9$ (solid black line) with fixed coupling ratio (f) $\lambda_{2}/\lambda_{1}=0$, (g) $\lambda_{1}/\lambda_{2}=0$, (h) $\lambda_{2}/\lambda_{1}=0.5$, (i) $\lambda_{1}/\lambda_{2}=0.5$ and (j) $\lambda_{2}/\lambda_{1}=1$. 
The other system parameters are given by $T_{h}=0.5$, $T_{c}=0.1$,
$\omega_{h} = 2\omega$, $\omega_{c} = \omega$. All quantities above are in units of $\omega$.
}\label{Stark10}
\end{figure*}

%==========================================

%==========================================
\begin{figure*}[tbp]
\begin{center}
%\vspace{-2.2cm}
\includegraphics[width=0.9\textwidth]{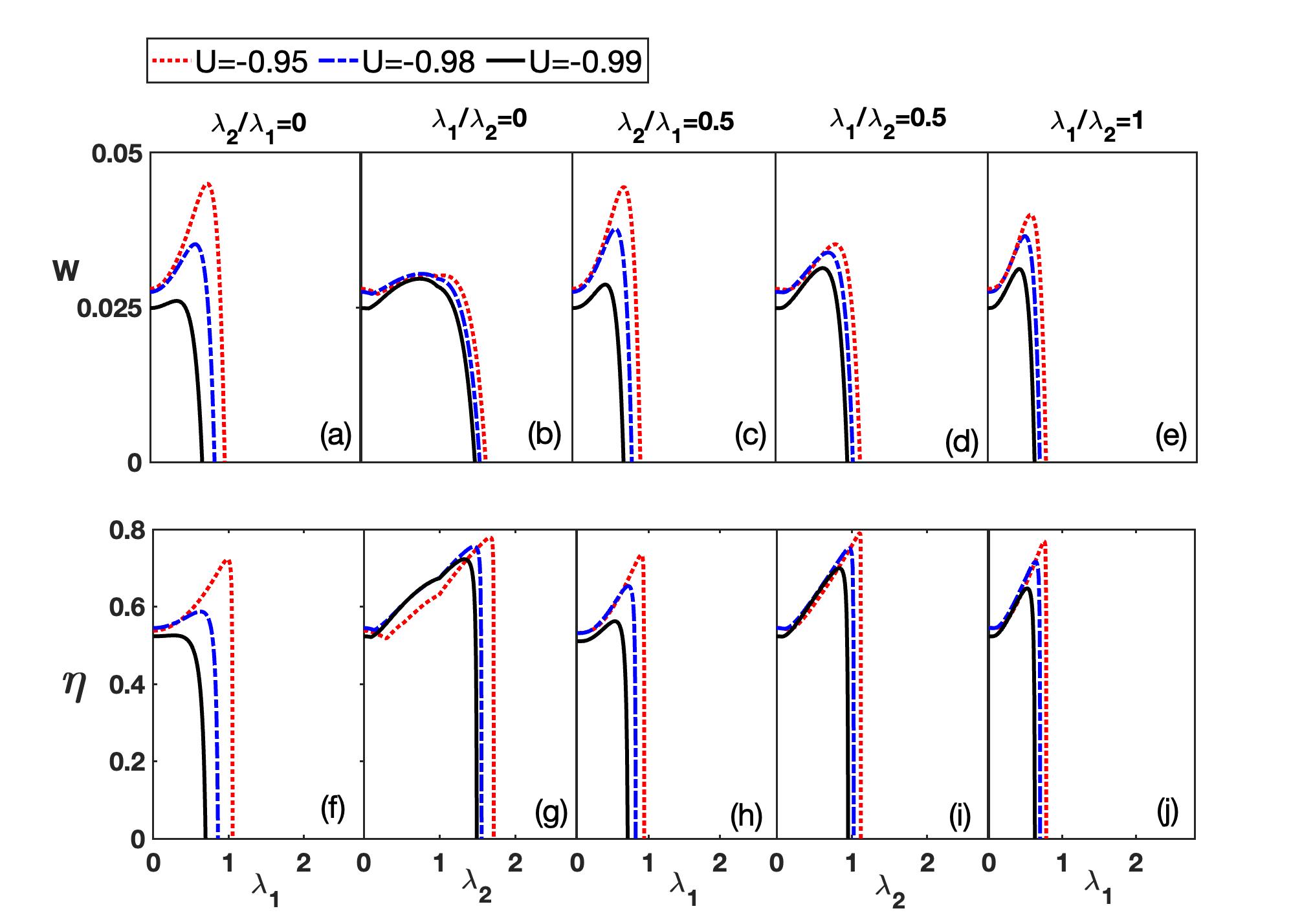}
%\vspace{-1.0cm}
\end{center}
\caption{Work W and  Efficiency $\eta$ as a function of the qubit-boson coupling strength ratio $\lambda_{1}/\lambda_{2}$ for several nonlinear Stark coupling strengths  $U=-0.95$ (dotted red line), $U=-0.98$ (dash dot  blue line), and $U=-0.99$ (solid black line). (a)-(e) Fixing (a) $\lambda_{2}/\lambda_{1}=0$, (b) $\lambda_{1}/\lambda_{2}=0$, (c) $\lambda_{2}/\lambda_{1}=0.5$, (d) $\lambda_{1}/\lambda_{2}=0.5$ and (e) $\lambda_{2}/\lambda_{1}=1$. (f)-(j) Fixing (f)$\lambda_{2}/\lambda_{1}=0$,(g)$\lambda_{1}/\lambda_{2}=0$,(h)$\lambda_{2}/\lambda_{1}=0.5$, (i)$\lambda_{1}/\lambda_{2}=0.5$ and (j)$\lambda_{2}/\lambda_{1}=1$. 
The other system parameters are given by $T_{h}=0.5$, $T_{c}=0.1$,
$\omega_{h} = 2\omega$, $\omega_{c} = \omega$. All quantities above are in units of $\omega$.
}\label{Stark11}
\end{figure*}
%==========================================

%==========================================
\begin{figure*}[!htbp]
\begin{center}
%\vspace{-2.2cm}
\includegraphics[width=0.9\textwidth]{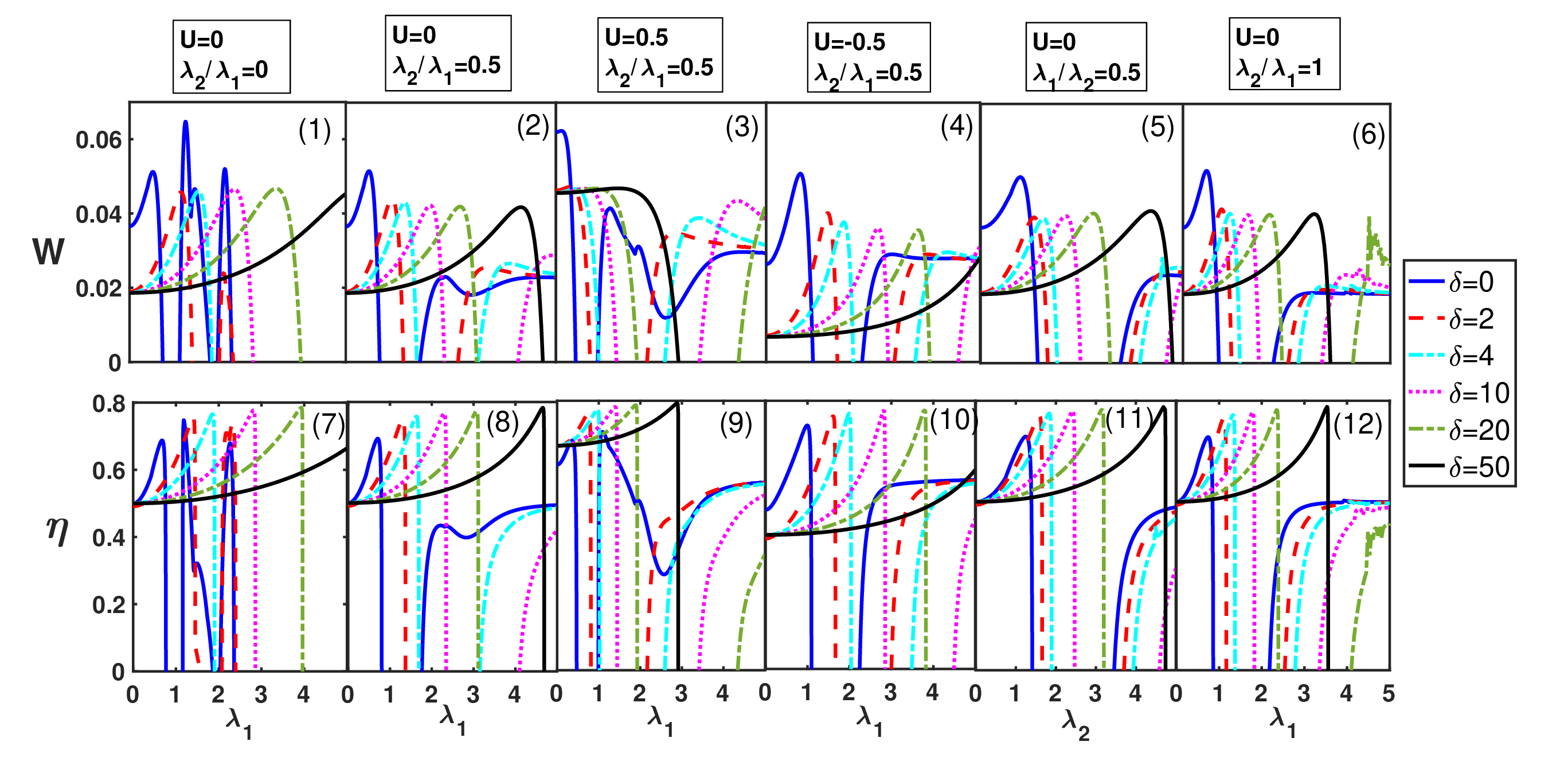}
%\vspace{-1.0cm}
\end{center}
\caption{(a)-(e) Panels (1) to (12) illustrate the Work (W) and Efficiency (\(\eta\)) as functions of the qubit-boson coupling strength \(\lambda_1\) (or \(\lambda_2\)) for different nonlinear Stark coupling strengths (U) and detuning (\(\delta\)). The top row (panels 1 to 6) shows the Work (W) and the bottom row (panels 7 to 12) shows the Efficiency (\(\eta\)). Each panel represents different values of U and coupling ratios (\(\lambda_2/\lambda_1\) or \(\lambda_1/\lambda_2\)), as indicated in the top of each panel. The curves within each panel represent different detuning values (\(\delta = 0, 2, 4, 10, 20, 50\)) as indicated in the legend. The temperatures are fixed at \(T_c = 0.1\) and \(T_h = 0.5\). Note that efficiency peaks with increasing detuning $\delta$ and decreases for negative values of $U$. All quantities are measured in units of \(\omega\).}
\label{detu}
\end{figure*}
%==========================================

%\begin{figure}[tbp]
%\begin{center}
%\vspace{-2.2cm}
%\includegraphics[width=0.5\textwidth]{Ottocorrelation.jpg}
%\vspace{-1.0cm}
%\end{center}
%\caption{The steady state Negativity and quantum discord of quantum Rabi-Stark model as a function of the qubit-boson coupling strength $\lambda_{1}$($\lambda_{2}$) (in units of $\omega$)for different nonlinear Stark coupling strength U with fix coupling ratio $\lambda_{2}/\lambda_{1}=0.5$ (the first row), $\lambda_{1}/\lambda_{2}=0.5$ (the second row), and $\lambda_{2}/\lambda_{1}=1$ (the third row). 
%The other system parameters are given by $T_{h}=0.5$, $T_{c}=0.1$,
%%$\omega_{h} = 2\omega$, $\omega_{c} = \omega$. All quantities above are in units of $\omega$.
%}\label{Stark0}
%\end{figure}

\section{Results and Discussions}

\subsection{Operational Modes for the Ideal Infinite Time Cycle.}
 The study of the ideal adiabatic regime (zero power) may lack direct practical relevance, yet it plays a crucial role in establishing the upper limits of thermodynamic figures of merit and offering a qualitative understanding of the different permissible operational regimes—Heat engine (E), Refrigerator (R), Heater (H), and Accelerator (A)—governed by Clausius's inequality and the first law of thermodynamics. 
 
 %Therefore, we initiate our investigation with this ideal scenario.
 
 %, which, to the best of our knowledge, remains largely unexplored in the existing literature.
 
We begin by examining the anisotropic model ($U = 0$, as illustrated in the operational phase diagram Fig.\ref{3DThTc} with $\omega_{h}/\omega_{c} = 2$, as a function of the coupling strength $\lambda_{1}(\lambda_{2})$ and the hot temperature of the reservoir $T_{h}$, for representative anisotropies values $r=0,1/2,1,2,\infty$ and cold temperatures of the reservoir $T_{c}=0.1,0.5,2$. By manipulating the parameter $\lambda$ within two given operating temperatures of the Otto cycle, we can achieve four distinct machine types. Notably, at low temperatures, Fig.\ref{3DThTc}(1)-(5), all four operational regions are present, predominantly occupied by the refrigerator (green area) and the heat engine (deep blue area). 

As anticipated, within the regions of the spectrum depicted in Fig.\ref{spectrum} that maintain harmonicity, all energy gaps undergo uniform changes during the quantum adiabatic process. Consequently, the positive working condition (PWC) mirrors the behavior observed in quantum harmonic oscillators or qubits, as described by $T_{h} > \frac{\omega_{h}}{\omega_{c}}T_{c}$ and confirmed in Fig.\ref{3DThTc}. As seen from the figure, the operational modes without practical applicability, such as the heater (yellow) and accelerator (red), emerge in regions where quasi-degenerate excited states are more pronounced, gradually diminishing as the temperature of the cold bath is increased.

Figures \ref{Ulambda}(a)-(h) depict the operational behaviors of the quantum Otto engine based on the nonlinear Stark coupling strength $U$ and qubit-boson coupling strength $\lambda_1$, with $r$ increasing from left to right across panels. Near critical points, abrupt transitions occur among heat engine, refrigerator, heater, and accelerator modes, where even small changes in parameters can significantly affect the engine's performance. The sensitivity to $\lambda_1$ and the range $-1 \leq U \leq 1$ emphasizes the profound impact of $U$ on these operational shifts. Lower cold reservoir temperatures stabilize the heat engine and refrigerator modes, whereas higher temperatures enable a broader range of operational modes. This highlights the complex dependence of the engine on coupling parameters and temperature, which is crucial for optimizing its performance.

The operational phase diagrams, Fig. \ref{3DThTc} and Fig. \ref{Ulambda}, illustrate the occurrence of unwanted heater and accelerator cycles around quasi-degenerate energy levels. This phenomenon is significant in both true first-order QPTs, as given by $\lambda_{1c}=\sqrt{\frac{\Delta(1-U^{2})}{U(1+r^{2})+1-r^{2}}}$, and in level degeneracies, such as those observed in the QRM.
In quantum engines, heater cycles involve energy absorption, while accelerator cycles involve rapid energy transfer, both of which are of no practical interest. It is noteworthy that close to the continuous QPT, the only region where the PWC is satisfied, leading to the Heat engine cycle, occurs when the Stark interaction $U$ is negative ($U \approx -1$), $\lambda_1 < 1$, and the system operates in a low-temperature regime ($T_c = 0.1$). For all other parameter configurations close to the second-order QPT, the engine can only function as a refrigerator.

\subsection{Ideal Infinite Time Cycle: Work and Efficiency.}
To conduct a more in-depth quantitative study of the model in question, in this work we focus on the heat engine regime of the AQRM working medium. Figure \ref{2DAQRM} presents computations of the net work $W$, Figs.\ref{2DAQRM}(a,d), the normalized work $\overline{W} = \frac{W}{W_{\mathrm{qubit}} + W_{\mathrm{QHO}}}$, Figs.\ref{2DAQRM}(b,e), and efficiency $\eta $, Figs.\ref{2DAQRM}(c,f), as functions of the coupling strengths $\lambda_1$ (RW) and $\lambda_2$ (CRW).
 The left column presents outcomes for temperatures $T_c = 0.1$ and $T_h = 0.5$, while the right column elucidates results for $T_c = 0.5$ and $T_h = 2$.

For reference, the qubit and quantum harmonic oscillator engine maintain a constant efficiency of $\eta = 0.5$, with work $W_{\mathrm{qubit}} \approx 0.018$ ($0.15$) and $W_{\mathrm{QHO}} \approx 0.019$ ($0.43$), corresponding to the parameters in the first (second) column of Fig.\ref{2DAQRM}. The primary advantage of AQRM over the uncoupled system in generating net work is evident in the low-temperature cycle regime of $T_c = 0.1$ and $T_h = 0.5$, as depicted in Fig.\ref{2DAQRM}(a). This advantage is particularly notable within the coupling region bounded by $1.2 \leq \lambda_1 \leq 1.7$ and $0 \leq \lambda_2 \leq 0.5$. In this region, the maximum value of $W_{\mathrm{max}} = 0.073$ is achieved for $\lambda_1 = 1.41$ and $\lambda_2 = 0.22$. Remarkably, $W_{\mathrm{max}}$ is twice the amount of work delivered by a decoupled qubit plus QHO heat engine. It's worth noting that in the deep strong coupling regime at low temperature, Fig.\ref{2DAQRM}(b), and for all coupling regimes at higher temperatures, Fig.\ref{2DAQRM}(e), it becomes apparent that the decoupled system can deliver more work during the engine cycle.
As observed in Figs.\ref{2DAQRM}(c),(f), for a broad range of couplings, $\eta$ surpasses the quantum Otto efficiency of a working medium with a harmonic spectrum, $\eta_{\alpha} = 1 - \omega_{c} / \omega_{h}=0.5$. Specifically, $\eta_{\mathrm{max}} = 0.75$ ($0.64$) for the low (high) temperature regime, very close to the Carnot efficiency, $\eta_{\text{Carnot}} = 0.8$ ($0.75$).

We now turn our attention to evaluating the unique effect of the temperature gradient on engine performance. To do so, we focus on the Rabi model ($r=1$) by computing the figures of merit, $W$, $\overline{W}$, and $\eta$, for fixed cold (hot) reservoir temperatures $T_c=0.1$ ($T_h=4$), varying the light-matter coupling $\lambda$ and the hot (cold) bath $T_h$ ($T_c$) temperature as depicted in Fig.\ref{2DAQRM2}, left (right) columns. In the scenario of a low-temperature bath at $T_c=0.1$, it's evident from $\overline{W}$ that the coupled system can surpass the total net work generated by the engine driven by the qubit and QHO working in parallel by more than $600\%$ in the temperature gradient region where $T_h<0.5$, and the coupling strength falls within $0.1 < \lambda < 1$. Additionally, within this parameter regime, $\eta_{\mathrm{max}}$ is approximately equal to $\eta_{\text{Carnot}}$. As depicted in Fig.\ref{2DAQRM2}(e), for nearly all parameters, the net work remains the same as that of the decoupled system, except in the deep strong coupling regime, $\lambda > 2$, and at higher cold bath temperatures, $T_c > 1.8$. Here, the PWC for the decoupled system, $T_{h} > \frac{\omega_{h}}{\omega_{c}} T_{c}$, is nearly saturated, resulting in almost negligible total work ($W_{\mathrm{qubit}} \approx W_{\mathrm{QHO}} \approx 0$), indicating a looser PWC for the QRM in this region. Surprisingly, as shown in Fig.\ref{2DAQRM2}(f), the coupled-system WM operates with higher efficiency across a wide range of parameters in high-temperature engine operation, where quantum correlations are absent\cite{xu2024persisting}.
Examining Figs. \ref{combined_figures}(a)-(f), we observe that the performance of a quantum Otto engine based on the ARSM varies significantly with temperature conditions, the coupling parameter $\lambda$, and the nonlinear Stark coupling $U$. Under conditions where $T_c = 0.1$ and $T_h = 0.5$, both the work $W$ and efficiency $\eta$ are maximized within specific ranges of $\lambda$ and $U$, approaching the Carnot efficiency $\eta_{Carnot} = 0.8$. The normalized work $\overline{W} > 1$ suggests that under these conditions, the engine can outperform the combined capabilities of qubit and quantum harmonic oscillator engines. Conversely, for $T_c = 0.5$ and $T_h = 2$, both the work and efficiency decrease as expected with higher temperatures. These observations underscore the critical role of temperature and coupling parameters in optimizing the engine's performance, emphasizing the need to explore different coupling regimes to maximize efficiency and work output.
Analyzing Figs. \ref{Stark9}(a)-(f), we find that the work extracted from the quantum heat engine depends on the qubit-boson coupling strengths $\lambda_1$ ($\lambda_2$) and the nonlinear Stark coupling strength $U$. Positive values of $U$ lead to an increase in work up to an optimal point before decreasing, while negative $U$ results in a consistent decrease in work. The work output is also markedly influenced by variations in $\lambda_1$ and $\lambda_2$, with the anisotropy ratio $\lambda_2/\lambda_1$ exhibiting diverse behaviors and indicating optimal operational points for the engine.
Achieving optimal performance in the engine requires precise adjustments of $U$, $\lambda_1$, and $\lambda_2$. Positive $U$ proves advantageous up to a certain limit, and the ratio between $\lambda_2$ and $\lambda_1$ plays a crucial role in maximizing performance. These findings underscore the complexity of optimizing quantum heat engines and highlight the importance of parameter exploration in achieving desired efficiencies.

To investigate the ideal quasi-static conditions and comprehend the variations in efficiency $\eta$ of a quantum heat engine as influenced by the coupling parameters ($\lambda_1$, $\lambda_2$) and the nonlinear Stark coupling strength $U$, Figs. \ref{Stark10}(a)--(j) illustrate how the efficiency approaches the Carnot limit and is impacted by proximity to critical points. The efficiency maximizes when $\lambda_1$ and $\lambda_2$ lie between 1.0 and 1.5, and for positive $U$ values ranging from 0.5 to 0.9, with a noticeable decline outside these ranges. Close to critical points connected to quantum phase transitions, an initial efficiency increase is triggered by level structure effects, followed by sudden decreases, rendering engine operation impracticable. Figs. \ref{Stark10}(a)--(e) indicate higher efficiencies with positive $U$, peaking near 0.75, whereas Figs. \ref{Stark10}(f)--(j) show lower efficiencies with negative $U$, often nearing Carnot efficiency for certain parameter sets. This study highlights the potential to enhance quantum heat engine efficiency by methodically choosing parameters, particularly near critical points.

Similarly, Figs. \ref{Stark11}(a)--(e) and (f)--(j) display the work $W$ and efficiency $\eta$ produced by the quantum heat engine as functions of the coupling parameters $\lambda_1$, $\lambda_2$, and the nonlinear Stark coupling strength $U$. These figures highlight the crucial impact of the coupling strengths and the Stark parameter on the work output near the second-order quantum phase transition. Due to the spectrum collapse, achieving convergence in numerical simulations near the critical point requires considerable computational resources. It is notable that the model is well-defined for all parameter values as the Hamiltonian operator remains self-adjoint, although it lacks a ground state above the critical point \cite{braak2024spectral}. Thus, our analysis focuses on three illustrative Stark coupling strengths.

Significant peaks in the work are observed, with the maximum work output declining as the system nears the critical point. Notably, Figs. \ref{Stark11}(a)--(e) show that around the critical point, work extraction is more sensitive to the anisotropy $r$ than in other regions, as seen in Fig. \ref{Stark9}, indicating the difficulty of tuning parameters for optimal engine performance. Conversely, Figs. \ref{Stark11}(f)--(j) indicate that the engine can achieve efficiencies very close to the Carnot limit for negative $U$, emphasizing the sensitivity of the engine to both the sign and magnitude of $U$. This thorough analysis highlights the essential role of coupling parameters and nonlinear interactions in optimizing the performance of quantum heat engines.

Additionally, analysis of Figs. \ref{detu}(1)--(12) suggests that under detuning conditions, the efficiency $\eta$ of the quantum heat engine can approach Carnot efficiency. This is depicted in Fig. \ref{detu}, where the efficiency peaks notably, particularly with increasing detuning $\delta$. Generally, negative values of $U$ decrease the efficiency. Achieving optimal efficiency requires fine-tuning parameters such as detuning $\delta$, the nonlinear Stark coupling strength $U$, and the coupling ratios $\lambda_2/\lambda_1$ and $\lambda_1/\lambda_2$. Therefore, under ideal conditions, the efficiency of the quantum heat engine can nearly reach Carnot efficiency at the cost of lower work output and deep-strong light-matter couplings, as seen in the figures. Crucially, controlling light-matter detuning offers a valuable mechanism to enhance the performance of quantum heat engines across various operational scenarios.

%==========================================
\begin{figure*}[tbp]
\begin{center}
%\vspace{-2.2cm}
\includegraphics[width=1\textwidth]{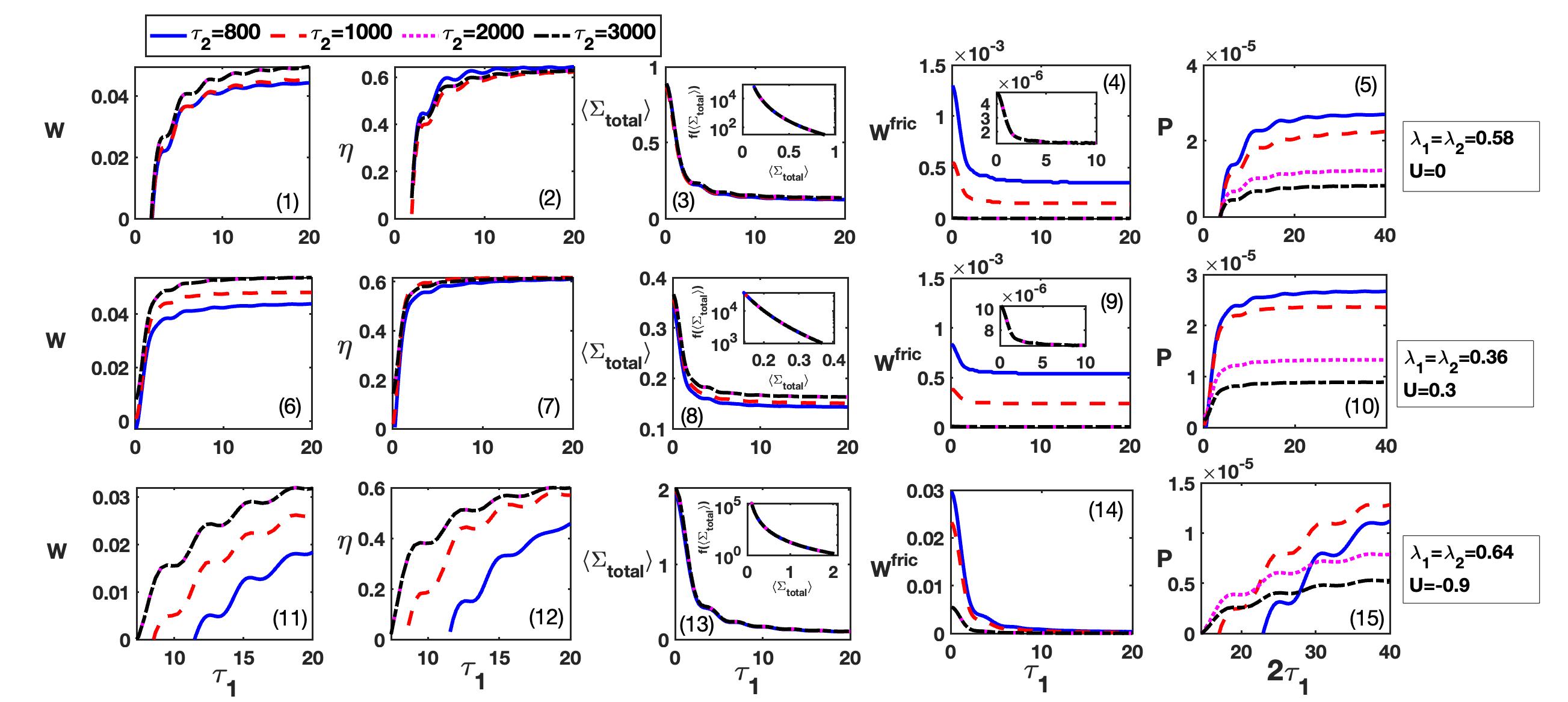}
%\vspace{-1.0cm}
\end{center}
\caption{Panels (1)-(15) illustrate, from left to right, the total work $W$, efficiency $\eta$, total entropy production $\langle \Sigma_{\text{total}} \rangle$, friction work $W^{\text{fric}}$, and power $P$ as functions of the adiabatic driving time $\tau_3 = \tau_1$ and thermalization stroke time $\tau_2 = \tau_4$, for various Rabi couplings $\lambda_1=\lambda_2$ and Stark strength $U$, as indicated in the rightmost column of all figures. The temperatures considered are $T_h = 0.5$ and $T_c = 0.1$. The subplots on total entropy production (3),(8) and (13) indicate the lower bound of the thermodynamic uncertainty relation with respect to $\langle \Sigma_{\text{total}} \rangle$. The subplots on total friction work (4) and (9) are zoomed-in views of the corresponding larger figures.
Additional cycle parameters include  $\omega_h = 2\omega$ and $\omega_c = \omega$, with all quantities expressed in units of $\omega$.}
\label{Stark13}
\end{figure*}
%==========================================
\subsection{Finite Time Cycle.}
Finite-time quantum heat engines hold significant promise for practical applications in quantum technologies, particularly in quantum information processing where rapid and efficient thermal cycling is essential. These engines also provide valuable insights into the fundamental limits of thermodynamic processes in the quantum regime, enhancing our understanding of how quantum effects impact thermodynamic performance.
In light of the findings from previous sections, we conduct an analysis focusing on three optimal parameter regimes near the critical points of the first- and second-order QPT as depicted in Fig. \ref{Stark13}, which displays the time-dependent thermodynamic quantifiers: work $W_{(t)}$, efficiency $\eta_{(t)}$ and power $P_{(t)}$. Operating under finite-time constraints, where cycle lengths are finite rather than infinitely long, poses several challenges that directly influence engine performance and efficiency. We evaluated the detrimental effects of finite-time processes in Fig. \ref{Stark13} through the total entropy $\left\langle \Sigma_{\text{total}} \right\rangle$ as defined in Eq. \ref{eq:entropy} and the work due to internal friction $W^{\text{fric}}$ calculated from Eq. \ref{eq:work}. We note that all values are calculated after the system reaches the limit cycle. The necessary number of cycles depends on the total cycle length and the coupling parameters. For this study, three unique sets of coupling parameters are chosen to thoroughly cover both quantitative and qualitative results across the whole parameter space. Additionally, the important topic of finite-time optimization will be reserved for further research
As illustrated in the first column of Fig. \ref{Stark13}, the total amount of work extracted diminishes for nonequilibrium states reached with short thermalization stroke times $\tau_{2} = \tau_{4}$ and brief adiabatic stroke times $\tau_{1} = \tau_{3}$. Interestingly, for couplings near the first-order critical point, as evident in Fig. \ref{Stark13}(1) and (6), a thermalization time of $\tau_{2} \approx 2000$ and adiabatic transitions with $\tau_{1} > 10$ are sufficient to recover the maximum work realized in the ideal case. On the other hand, close to the continuous QPT, achieving the same work as in the infinite time cycle is difficult and not achievable within the simulation times considered, as demonstrated in Fig. \ref{Stark13}(11). Similarly, the efficiency of the system also decreases for nonequilibrium states with short thermalization and adiabatic stroke times, as shown in Fig. \ref{Stark13}(2),(7) and (12). Particularly around the first-order critical point, the efficiency reaches the ideal maximum value with a thermalization time of $\tau_{2} \approx 800$ and adiabatic changes with $\tau_{1} > 5$, matching the behavior of the extracted work. Nonetheless, near the continuous QPT, achieving the same level of efficiency as in the infinite time cycle is equally challenging and impossible within the simulation times considered.
The performance of quantum engines is critically dependent on power output, with practical applications often necessitating substantial power delivery within short durations. This significance is demonstrated in the final panel of Fig. \ref{Stark13}. The findings indicate that, unlike work and efficiency, power is less sensitive to the thermalization time, resulting in comparable power outputs for nonequilibrium states with brief thermalization periods near the first-order critical point, as depicted in Fig. \ref{Stark13}(5) and (10). Notably, near the continuous QPT, achieving peak power output is as challenging as attaining optimal work and efficiency, rendering it unattainable within the simulated times, as illustrated in Fig. \ref{Stark13}(15), where the generated power is considerably lower by an order of magnitude compared to other cases.

The importance of power in finite-time quantum engines extends to various applications, ensuring optimal system performance within finite operational durations, thereby enhancing overall practicality and effectiveness. Moreover, in the development of nanoscale heat engines, where achieving high efficiency and power in brief cycles is challenging, understanding the trade-offs between stroke times and performance metrics is crucial for advancing technology in energy conversion and management at the quantum level. Therefore, investigating the conditions under which maximum power can be achieved, particularly near critical points, provides valuable insights for designing more efficient and powerful quantum devices.

To deepen our understanding of the finite-time thermodynamic quantifiers mentioned earlier, we evaluate the adverse effects of finite-time operation through total entropy $\left\langle \Sigma_{\text{total}} \right\rangle$, as illustrated in Figs. \ref{Stark13}(3), (8), and (13). The subplots show the computations of the lower bound of thermodynamic uncertainty relations as defined in Eq. \ref{Timpa}, and the impact of quantum nonadiabaticity in evolution through friction work ($W^{\text{fric}}$), shown in Figs. \ref{Stark13}(4), (9), and (14).

From Figs. \ref{Stark13}(3),(8),(13) we observe that the total entropy changes with thermalization and adiabatic times are too small to be noticeable on the scale where the entropy is changing for all graphs. The uncertainty relation shows that for a small total entropy reached in the longer time limit, the power or efficiency must tend to zero, or the variance in this quantity will go to infinity. Conversely, when the thermal quantizer is maximized for short adiabatic times, the AQRSM around the first-order QPT can lead to small fluctuations, while the second-order point results in large signal-to-noise ratios. Let us recall that Eq. \ref{Timpa} provides the best case for the efficiency and power of the stochastic cycle, and a detailed study of the quantum fluctuations of the AQRSM as a working fluid is in demand and will be explored in a future work. 

%is beyond the scope of our study.

Furthermore, for the friction work plots, Figs. \ref{Stark13}(4),(9),(14) we observe that although the total power around the first-order QPT shows a minor dependence on the thermalization time, the amount of friction changes by three orders of magnitude between the fast and slow thermalization strokes. As illustrated in the figure, the friction work demonstrates a strong dependence on the adiabatic and thermalization times around the second-order QPT point, leading to increased wasted work due to friction in this scenario.

We note that to mitigate the effects of quantum friction and irreversibility, several optimization strategies can be employed. One such approach is shortcut-to-adiabaticity (STA) protocols, which are designed to mimic finite-time adiabatic transformations \cite{Torrontegui2013, Deffner2014}. These protocols use contradiabatic driving or other techniques to suppress nonadiabatic excitations, thereby reducing quantum friction and increasing efficiency \cite{delCampo2013}. Another strategy  focuses on fine-tuning  control parameters and  cycle timings to balance both power output and efficiency \cite{Abah2018}. This form of optimization necessitates a thorough comprehension of the system's dynamics and the interaction between various dissipation mechanisms \cite{Niedenzu2019}. These strategies are essential in crafting quantum thermodynamic cycles, aiming to achieve high efficiency while controlling dissipation \cite{Gelbwaser-Klimovsky2016}. We anticipate that incorporating these techniques in subsequent research will enhance the efficiency of quantum heat engines operating within finite-time frameworks near critical points.

\section{Conclusion}
In this study, we thoroughly examined the functionality of a quantum Otto heat engine using the anisotropic quantum Rabi-Stark model (AQRSM) as the working medium. Our research highlighted that quantum phase transitions, which influence the spectral characteristics of the medium—such as level crossings, degeneracies, and quasi-degenerate points—play a crucial role in optimizing the total work output and efficiency of the quantum Otto cycle.

In particular, the observed first-order and continuous phase transitions in the AQRSM provided significant insights into leveraging critical properties to enhance the performance of quantum heat engines. In the theoretical scenario of infinite operation time with zero power output, we constructed a detailed phase diagram delineating the operational states of the machine, specifying the conditions under which it functions as a heat engine, refrigerator, heater, or accelerator.

Furthermore, by considering finite-time operations, we investigated the effect of quantum friction and conducted a limit cycle analysis, underscoring the importance of the spectral properties of the working substance in determining the optimal operating conditions for the quantum Otto engine. Our study indicates that quantum heat engines based on coupled critical systems have the potential to significantly surpass the performance of their isolated counterparts working independently. This finding opens promising avenues for advancements in renewable energy technologies and quantum computing, particularly in cryogenically cooled environments that require efficient energy and thermal management.

We advocate for future research to focus on the practical implementation of the AQRSM-based thermal engine in accessible physical systems and to examine the statistical outcomes derived from stochastic thermodynamic processes.

\section{Acknowledgement}
We acknowledge financial support from the Brazilian agencies CNPq and FAPEG. This work was performed as part of the Brazilian National Institute of Science and Technology (INCT) for Quantum Information Grant No. 465469/2014-0.
H.-G. X. and J. J. are supported by National Natural Science Foundation of China under Grant No. 11975064.

\appendix
\renewcommand\thesection{\Alph{section}} \renewcommand{\thefigure}{\thesection.\arabic{figure}}
\setcounter{equation}{0}
\setcounter{figure}{0}
\setcounter{table}{0}

%\subsection{Approximation of the two-photon correlation function}
%The two-photon correlation function is defined at Eq.(\ref{gn2}) as
%$G^{(2)}_N(0)=\frac{{\langle}(\hat{X}^-)^2(\hat{X}^+)^2{\rangle}}{{\langle}\hat{X}^-\hat{X}^+{\rangle}^2}$,
%with $N$ the qubits number and
%$\hat{X}^+=-i\sum_{k>j}\Delta_{kj}X_{jk}|\phi_j{\rangle}{\langle}\phi_k|$.
%For the term ${\langle}\hat{X}^-\hat{X}^+{\rangle}$, it can be expressed as
%\begin{eqnarray}
%{\langle}\hat{X}^-\hat{X}^+{\rangle}=\sum^{\infty}_{n=1}P_n\mathcal{A}_n,
%\end{eqnarray}
%where $P_n$ is the steady state population and
%the transition coefficient is $\mathcal{A}_n=\sum_{k<n}(\Delta_{nk}X_{nk})^2$.
%While the other term ${\langle}(\hat{X}^-)^2(\hat{X}^+)^2{\rangle}$ is expressed as
%\begin{eqnarray}
%{\langle}(\hat{X}^-)^2(\hat{X}^+)^2{\rangle}=\sum^{\infty}_{n=1}P_n\mathcal{B}_n
%\end{eqnarray}
%with the coefficient
%$\mathcal{B}_n=\sum_{k<l<n}(\Delta_{nk}\Delta_{kl}X_{nk}X_{kl})^2$.

%In the low temperatures regime, the system dynamics is mainly described by the several lowest energy levels due to weak thermal excitation.
%Hence, these two correlation function may be approximately expressed as
%\begin{eqnarray}
%{\langle}\hat{X}^-\hat{X}^+{\rangle}&=&P_1\mathcal{A}_1+P_2\mathcal{A}_2+P_3\mathcal{A}_3,\\
%{\langle}(\hat{X}^-)^2(\hat{X}^+)^2{\rangle}&=&P_2\mathcal{B}_2+P_3\mathcal{B}_3+P_4\mathcal{B}_4.
%\end{eqnarray}

\newpage
\bibliographystyle{unsrt}
\bibliography{bibliography.bib}

\end{document}